\documentclass[twocolumn]{aastex62}

\usepackage{color}
\usepackage{soul} 

\newcommand{\lenstool}{{\tt{Lenstool}}}

\newcommand{\hst}{{\it HST}}
\newcommand{\HST}{{\it HST}}
\newcommand{\Lya}{Ly$\alpha$}

\newcommand{\Halpha}{H$\alpha$}
\newcommand{\Hgamma}{H$\gamma$}

\newcommand{\OIII}{$\lambda\lambda$4960,5008}
\newcommand{\OII}{$\lambda\lambda$3727,3729}
\newcommand{\kms}{km.s$^{-1}$}
\newcommand{\msun}{M$_{\odot}$}
\newcommand{\clusterfullname}{SPT-CL\,J0356$-$5337}
\newcommand{\clustername}{SPT$-$0356}
\newcommand{\clusterz}{1.0359} 
\newcommand{\zspecA}{2.363} 
\newcommand{\zspecB}{2.364} 
\newcommand{\zspecC}{3.048} 
\newcommand{\MSPTnew}{M$_{500 c}=3.59^{+0.59}_{-0.66} \times10^{14}h^{-1}_{70}$\, \msun} 

\newcommand{\MBocquetfixed}{M$_{500 c}=3.59 ^{+0.59}_{-0.66} \times10^{14}h^{-1}_{70}$\, \msun} 

\newcommand{\Mfivekpc}{M$_{500~{\rm kpc}}=4.0 \pm 0.8\times10^{14}\, $\msun\,} 
\newcommand{\Erad}{$\theta_{E}\simeq$\, 14\arcsec\ } 
\newcommand{\CMR}{$75\farcs6$}

\graphicspath{{./}{figures/}}

\received{---}
\revised{---}
\accepted{---}
\submitjournal{ApJ}

\shorttitle{Strong lensing analysis of a major merger at $z\sim$1}
\shortauthors{Mahler et al.}

\begin{document}

\title{Strong Lensing Model of \clusterfullname, a Major Merger Candidate at Redshift \clusterz}

\correspondingauthor{Guillaume Mahler}
\email{gmahler@umich.edu}

\author[0000-0003-3266-2001]{Guillaume Mahler}
\affil{Department of Astronomy, University of Michigan, 1085 South University Ave, Ann Arbor, MI 48109, USA}

\author[0000-0002-7559-0864]{Keren Sharon}
\affiliation{Department of Astronomy, University of Michigan, 1085 South University Ave, Ann Arbor, MI 48109, USA}

\author[0000-0003-1370-5010]{Michael D. Gladders}
\affiliation{Department of Astronomy and Astrophysics, University of Chicago, 5640 South Ellis Avenue, Chicago, IL 60637, USA}
\affiliation{Kavli Institute for Cosmological Physics, University of Chicago, 5640 South Ellis Avenue, Chicago, IL 60637, USA.}

\author[0000-0001-7665-5079]{Lindsey Bleem}
\affiliation{Kavli Institute for Cosmological Physics, University of Chicago, 5640 South Ellis Avenue, Chicago, IL 60637, USA.}
\affiliation{Argonne National Laboratory, High-Energy Physics Division, 9700 S. Cass Avenue, Argonne, IL 60439, USA }

\author[0000-0003-1074-4807]{Matthew B. Bayliss}
\affil{Department of Physics, University of Cincinnati, Cincinnati, OH 45221, USA}
\affil{Kavli Institute for Astrophysics and Space Research, Massachusetts Institute of Technology,\\
77 Massachusetts Avenue, Cambridge, MA 02139, USA}

\author[0000-0002-2238-2105]{Michael S. Calzadilla}
\affil{Kavli Institute for Astrophysics and Space Research, Massachusetts Institute of Technology,\\
77 Massachusetts Avenue, Cambridge, MA 02139, USA}

\author[0000-0003-4175-571X]{Benjamin Floyd}
\affiliation{Department of Physics and Astronomy, University of Missouri, 5110 Rockhill Road, Kansas City, MO 64110, USA}

\author[0000-0002-3475-7648]{Gourav Khullar}
\affiliation{Department of Astronomy and Astrophysics, University of Chicago, 5640 South Ellis Avenue, Chicago, IL 60637, USA}
\affiliation{Kavli Institute for Cosmological Physics, University of Chicago, 5640 South Ellis Avenue, Chicago, IL 60637, USA.}

\author[0000-0001-5226-8349]{Michael McDonald}
\affil{Kavli Institute for Astrophysics and Space Research, Massachusetts Institute of Technology,\\
77 Massachusetts Avenue, Cambridge, MA 02139, USA}

\author[0000-0002-7868-9827]{Juan D. Remolina Gonz\'alez}
\affil{Department of Astronomy, University of Michigan, 1085 South University Ave, Ann Arbor, MI 48109, USA}

\author[0000-0002-6987-7834]{Tim Schrabback}
\affil{Argelander-Institut f\"{u}r Astronomie, Universit\"{a}t Bonn, Auf dem
H\"{u}gel 71, 53121, Bonn, Germany}

\author[0000-0002-2718-9996]{Antony A. Stark}
\affil{Center for Astrophysics $\mid$ Harvard \&\ Smithsonian, 60 Garden Street, Cambridge, MA 02138, USA}

\author{Jan Luca van den Busch}
\affil{Argelander-Institut f\"{u}r Astronomie, Universit\"{a}t Bonn, Auf dem
H\"{u}gel 71, 53121, Bonn, Germany}
\affil{Astronomisches Institut, Ruhr-Universit\"{a}t Bochum, Universit\"{a}tsstr 150, D-44801 Bochum, Germany}

\begin{abstract}
We present an analysis of the mass distribution inferred from strong lensing by \clusterfullname, a cluster of galaxies at redshift $z=\clusterz$ revealed in the follow-up of the SPT-SZ clusters. The cluster has an Einstein radius of \Erad for a source at $z=3$ and a mass within 500 kpc of \Mfivekpc. Our spectroscopic identification of three multiply-imaged systems ($z=\zspecA$, $z=\zspecB$, and $z=\zspecC$), combined with \HST\ F606W-band imaging allows us to build a strong lensing model for this cluster with an rms of $\leq0\farcs3$ between the predicted and measured positions of the multiple images. Our modeling reveals a two-component mass distribution 
in the cluster. One mass component is dominated by the brightest cluster galaxy and the other component, separated by $\sim$170 kpc, contains a group of eight red elliptical galaxies confined in a $\sim$9\arcsec\ ($\sim$70 kpc) diameter circle. 
We estimate the mass ratio between the two components to be between 1:1.25 and 1:1.58.
In addition, spectroscopic data reveal that these two near-equal mass cores have only a small velocity difference
of $\sim300$ \kms\ between the two components. This small radial velocity difference suggests that most of the relative velocity takes place in the plane of the sky, and implies that \clusterfullname\ is a major merger with
a small impact parameter seen face-on. 
We also assess the relative contributions of galaxy-scale halos to the overall mass of the core of the cluster and find that within 800 kpc from the brightest cluster galaxy about 27\% of the total mass can be attributed to visible and dark matter associated with galaxies, whereas only 73\% of the total mass in the core comes from cluster-scale dark matter halos. 

\end{abstract}

\keywords{gravitational lensing: strong - galaxies: clusters: individual: \clusterfullname}

\section{Introduction}
\label{sec:intro}

Embedded in the largest gravitationally-bound dark matter halos in the cosmic web, clusters of galaxies are excellent probes of the high-mass end of large scale structure formation. 
Models of hierarchical growth predict that the majority of the mass of a cluster halo accumulates through multiple minor-merger events, in which small galaxy-scale or group-scale halos fall into the cluster core. Major mergers (1:3) are rarer; statistically, a typical cluster-scale halo of M$_{200}\sim10^{14}$\msun\ at $z=0$ will have undergone one major-merger event throughout its evolution \citep{Fakhouri2008}. M$_{\delta}$ goes the mass in the radius R$_{\delta}$ that would be reach ${\delta}$ times the critical density of the universe at that redshifts.

Major mergers are also uniquely useful for studying the nature of dark matter (DM). For example, the separation between DM and gas in the Bullet Cluster (1E~0657$-$558) provides empirical evidence that favors cold dark matter over theories of modified gravity \citep{Clowe2006}. Analyses of mergers can also constrain the DM self-interaction cross-section (e.g. \citealt{Markevitch2004,Harvey2015}) and the large-scale matter-antimatter ratio \citep{Steigman2008}. 

Structure growth and mergers are studied in simulations and observed up to $z\sim2$.
\citet{McDonald2017} consider the density profiles of clusters out to redshift $\sim1.7$ that have been re-scaled to their R$_{500}$ radius, taking into account the critical density at each epoch, and find that outside the cluster cores the profiles are remarkably similar.
Cluster cores deviate from this self-similarity; the complexity of cluster cores can be well probed with a multi-wavelength/multi-scale 
approach, in particular, by including strong lensing analysis. Indeed, the angular extent of strong lensing features in cluster fields, from a few to a hundred arcseconds, corresponds to the scale of the cluster core -- a few to hundreds of kpc in projection.

Since the prototypical Bullet Cluster \citep{Markevitch2002} was first identified, a small number of other clusters with structure indicative 
of major mergers have been observed, and showed spatial dissociation of gas,
DM, and galaxies. Most of these systems are at low redshifts. Notable higher redshift systems are ``El Gordo'' (ACT-CL~J0102-4915) at $z$=0.870 \citep{Marriage2011,Menanteau2012,Zitrin2015,Cerny2018}, and the structure of CLJ0152-1347 at $z$=0.830 \citep{Massardi2010}, a complex system with two main subclusters separated by 722 kpc, complicated by at least one further merging subgroup.

Only a subset of mergers enable investigation of the full range of phenomena seen in bullet-like mergers, targeting the nature of DM \citep{Dawson2012}: such mergers are those that (1) occur between two subclusters of comparable mass, (2) have a small impact parameter, (3) are observed during the
short period when the cluster gas is significantly offset from the galaxies and DM, and (4) occur mostly transverse to the line of sight such that the apparent angular separation of the cluster gas from the galaxies and DM is maximized. 

In this paper we confirm the identification of \clusterfullname\ (hereafter \clustername) as a strong lensing cluster, report on spectroscopic measurements of redshifts of three lensed galaxies behind the cluster, and present the first strong lensing model of the cluster core. We argue that the observed properties of the cluster, combined with the strong lensing mass model, promotes \clustername\ as the highest-redshift major-merger cluster candidate.

At $z=\clusterz$, \clustername is one of the most distant clusters known with spectroscopically-confirmed strong lensing evidence at the cluster scale from multiple systems. The lensing geometry offers a unique opportunity to weigh the mass within the core of the cluster. 
Strong lensing clusters at redshifts $z\geq1$, include SPT-CLJ2011$-$5228 at $z = 1.06$ \citep{Collett2017}, which has only one multiply-imaged system; SPT-CLJ0546$-$5345
at $z=1.066$ \citep{Brodwin2010}, which shows evidence of strong lensing features \citep{Staniszewski2009}; and SPT-CLJ0205$-$5829 \citep{Stalder2013,Bleem2015} at $z = 1.322$ which has one arc with no published redshift. \cite{Wong2014} also report on a lensed galaxy with spectroscopic redshifts behind a cluster at $z=1.62$. However, the lensing signal comes essentially from the brightest cluster galaxy (BCG) embedded in the cluster, thus this lens offers little or no leverage on the cluster-scale mass distribution. The highest redshift cluster with strong lensing evidence currently published is IDCSJ1426 at $z=1.75$ \citep{Gonzalez2012}, but there is no public spectroscopic redshift measurement for the only giant arc reported. 

This paper is organized as follows: In section~2, we report on the identification and previous analyses of \clustername. In section~3, we describe the data that are used in this paper. In section~4, we define the cluster-member selection, which is an important input to the strong lens modeling described and analyzed in section~5. In section~6, we discuss our results and we summarize this work in section 7.

Throughout this paper, we adopt a standard $\Lambda$-CDM cosmology with $\Omega_m =0.3$, $\Omega_{\Lambda}=0.7$, and $h=0.7$. All magnitudes are given in the AB system \citep{Oke1974}.

\section{\clusterfullname}\label{sec:cluster}

\citet{Bleem2015} first identified and published \clustername\
as a strong-lensing cluster, as part of a catalog of galaxy clusters selected from South Pole Telescope (SPT) data based on the Sunyaev-Zel'dovich effect (SZ,\citealt{Sunyaev1970}). \citet{Bocquet2019} published an updated mass for this system of \MSPTnew\ assuming the same fixed $\Lambda$-CDM cosmology we adopt in this work.

\begin{figure*}
     \includegraphics[trim={ 0 2cm 0 0},clip,width=\textwidth]{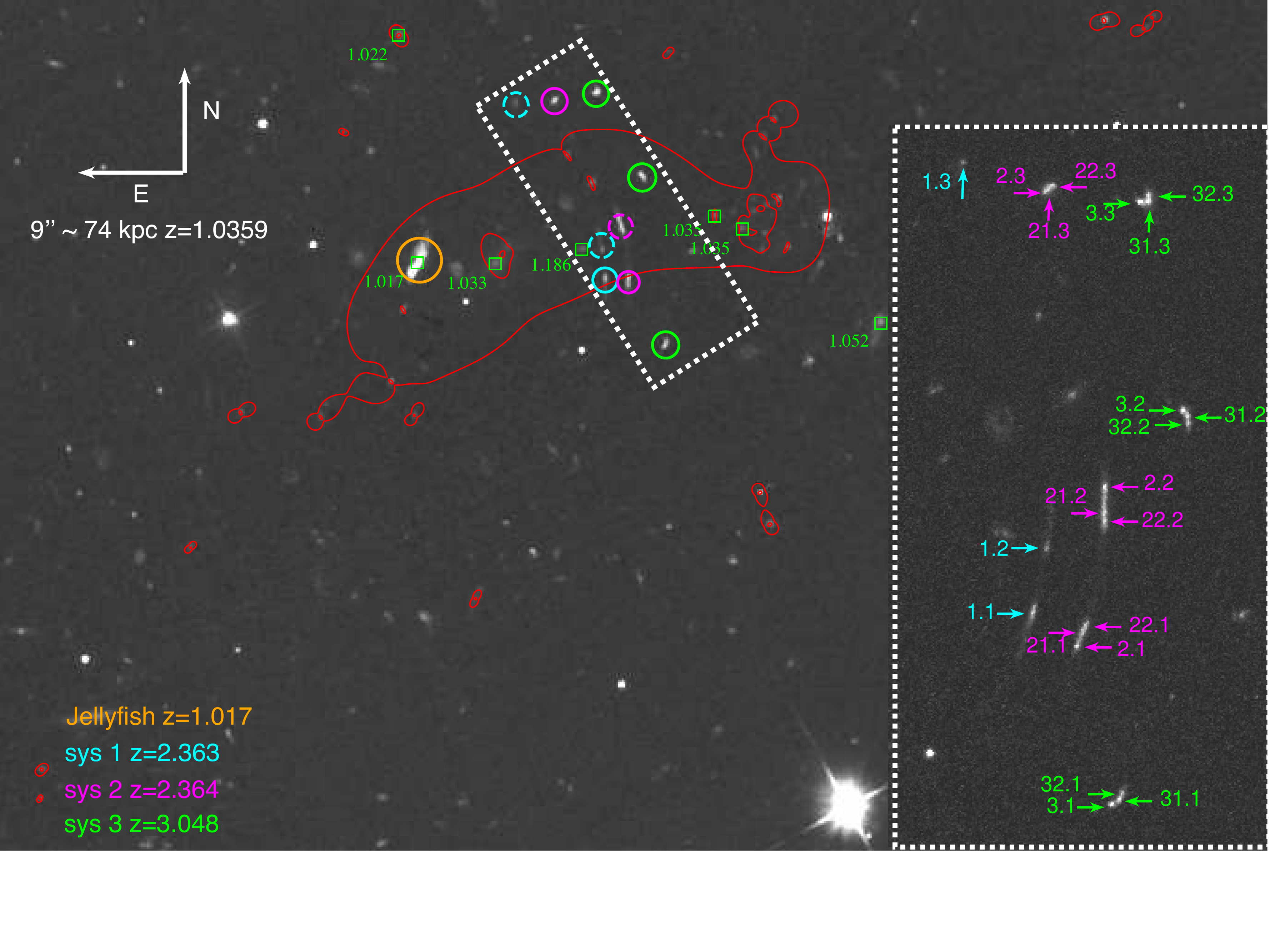}
         \caption{\HST/ACS imaging with the F606W filter, from GO-13412. 
         The red contours corresponds to the critical curve at redshift $z=3.048$ and matches the lens model with 1 DM component free to move, model B. (see sect.\,\ref{sec:lensing} for more details). The green squares label the GMOS spectroscopic data within the FoV from \citet{Bayliss2016}. All the other colored solid circles represent the FIRE data and match the colored legend in the figure. The dashed circles are the identified counter images of lensed systems. The white dashed box shows the position of the inset on the right. This inset shows a zoomed view of the three multiply-imaged systems, where the arrow and numbers indicate the lensing configurations and constraints used in the model following a color coding to match the lensed system.
}
     \label{fig:cluster}
 \end{figure*}

\citet{Bayliss2016} used the Gemini Multi-Object Spectrograph (GMOS) on the Gemini South Observatory in Chile to measure spectroscopic redshifts of 36 galaxies in this field, eight of which were spectroscopically identified as cluster members, including the BCG (see Figure \ref{fig:cluster}). From these eight cluster members with GMOS spectra, \citet{Bayliss2016}  determine a median cluster redshift of $z=1.0345\pm0.0112$, with a velocity dispersion $\sigma_{v} = 1691\pm588$ \kms. In a reanalysis of these data, \citet{Bayliss2017} report a revised median redshift of $z=1.0359\pm0.0042$ and $\sigma_{v} = 1647\pm514$ \kms, based on four of the eight galaxies whose spectral features indicate that they are either passive or post-starburst, so that their
velocities are likely less sensitive to recent accretion.
In this paper, we adopt as the cluster redshift the measurement of \citet{Bayliss2017}, $z=$ \clusterz, used hereafter without uncertainties. We note that these measurements are consistent with each other, and the slight difference between these redshifts has no significant effect on our analysis or results. 

\begin{center}
\begin{figure}
   \includegraphics[scale=0.55]{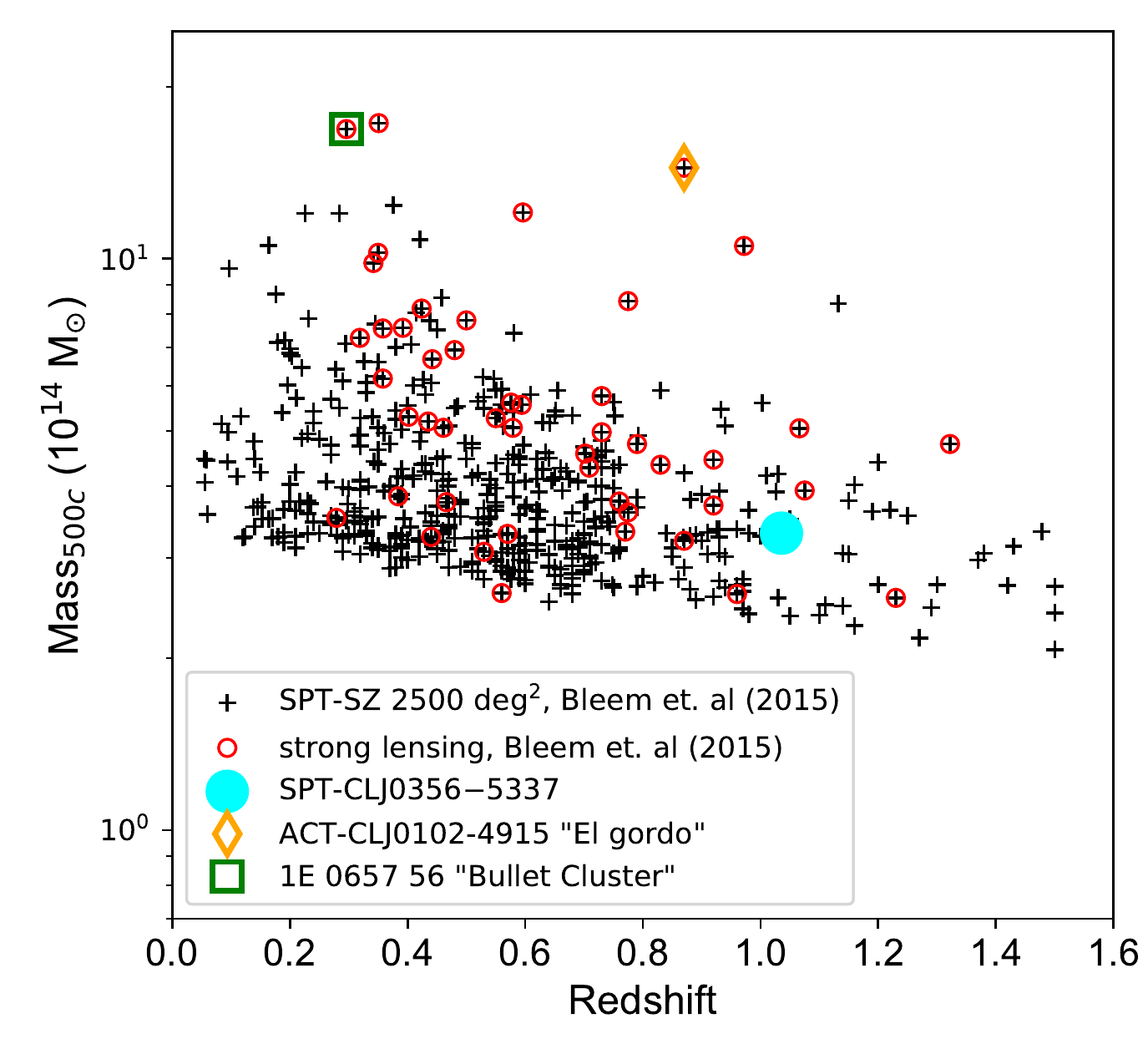}
        \caption{Comparison of \clustername\ to other clusters in the SPT-SZ 2500 deg$^2$ from \citet{Bleem2015}. Clusters identified as strong lenses are labeled with red circles, and several well-studied ``bullet'' clusters (i.e. dissociative merger) are highlighted. 
        \clustername\ is among the highest redshift strong lensing clusters in this sample.}
    \label{fig:SZcluster}
\end{figure}
\end{center}

Initial follow-up imaging of the SPT-SZ clusters led to the identification of strong lensing evidence in 23 clusters above $z>0.7$ \citep{Bleem2015}.
Figure~\ref{fig:SZcluster} plots the mass and redshift of \clustername\ compared to the entire \citet{Bleem2015} sample, with the strong lenses highlighted; \clustername\ is among the highest-redshift strong lenses in this sample.  
The lensing evidence, shown in Figure~\ref{fig:cluster}, include three sets of multiple images of background sources. Each system has three multiple images, all appearing West of the BCG. 
The high resolution
of the \HST/ACS single-band imaging (in F606W, shown in Figure~\ref{fig:cluster} ) revealed substructure in each of the
images, strongly suggesting that the three images
of each system are indeed multiple images of the same
source. This identification was also supported by the observed
symmetry, which is consistent with expectations
from the lensing geometry, and prompted follow-up spectroscopy. Here, we report nine lensed images of three distinct sources in the field of \clustername



\section{Data} \label{sec:data}

\subsection{Imaging}\label{sec:imaging}


Optical imaging follow-up observations of \clustername\ were conducted with several telescopes and instruments:
\paragraph{Magellan} 
The cluster was first imaged with the Inamori-Magellan Areal Camera \& Spectrograph (IMACS) on Magellan Baade 6.5-m telescope as part of the SPT cluster confirmation efforts on 2012 Dec 16. Each IMACS observation covers a field of $13'\times27'$, observed for 400~s with each of the $g$, $r$, and $z$ filters. 

The cluster was observed with the Magellan Clay 6.5-m telescope at Las Campanas Observatory, using the Parallel Imager for Southern Cosmology Observations (PISCO) instrument \citep{Stalder2014} as part of a uniform optical follow-up program on 2016 Dec 31.
Each PISCO observation covers a $9.5'\times6'$ area on the sky centered on the cluster, observed in parallel in four different bands ($g$, $r$, $i$, and $z$) for an exposure time of 258 seconds. 


\paragraph{Hubble Space Telescope}\HST\ imaging of \clustername\ was obtained with the Advanced Camera for Survey (ACS) camera as part of the SPT-SZ ACS Snapshot Survey
(Cycle 21, GO-13412; PI: Schrabback). A single image was obtained in F606W on 2014 June 25, with total exposure time of 2320~s. The ACS field of view covers a $3.3'\times3.3'$ area, centered on the SZ peak.

\paragraph{Gemini} Deep $i$-band and $g$-band images were obtained with the Gemini South Observatory 8.1-m telescopes as part of the weak lensing follow-up of SPT-SZ clusters (PI: Benson) from the SPT-SZ ACS Snapshot Survey using the GMOS camera. We used a $2\times2$ binned stacked image reaching 5200~s exposure time and 1\farcs17 seeing on the central chip covering the \HST/ACS field-of-view. The data were obtained on 2014 Dec 30.

\begin{figure*}
\begin{center}
   \includegraphics[scale=0.45]{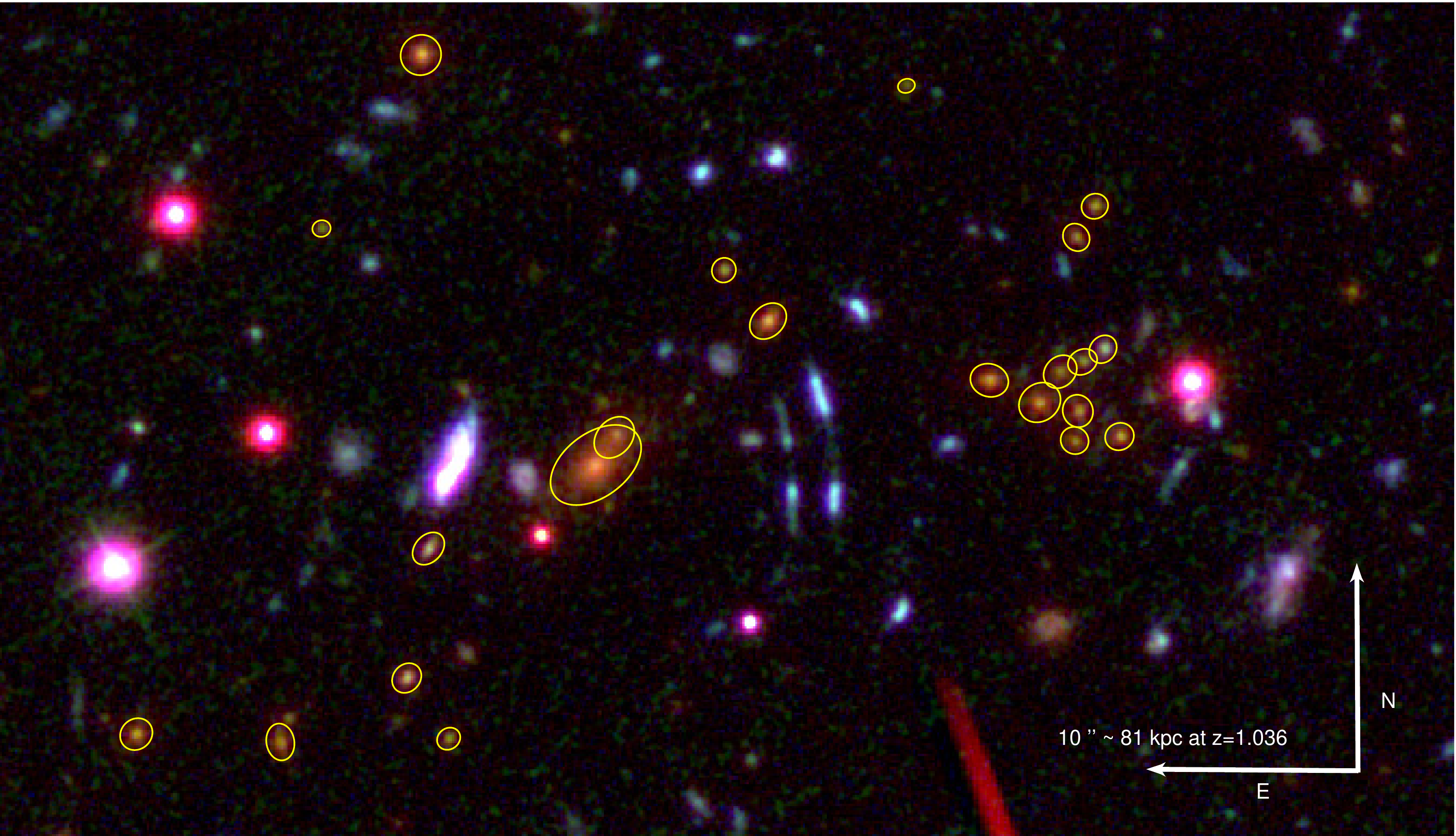}
\end{center}
        \caption{
        Color composite image of the field, rendered from Gemini $i$-band (red), \hst/ACS F606W (green) and Gemini 
        $g$-band (blue). The yellow ellipses mark selected cluster member galaxies that are in the field of view (see section \ref{sec:members})
        }
    \label{fig:PISCO}
\end{figure*}

Figure \ref{fig:PISCO} shows the cluster core field of view, rendered from the GMOS and \HST\ imaging. These represent the deepest and highest resolution data in hand. We supplement these data with the shallower, lower resolution, $z$-band imaging from  IMACS and PISCO, in order to achieve broad wavelength coverage, where color information is needed for assessing candidate strong lensing features. 



\subsection{Spectroscopy of Lensed Sources}\label{sec:spectroscopy}
\paragraph{Gemini/GMOS} The Gemini/GMOS-South spectroscopic survey of SPT-SZ clusters \citep{Bayliss2016} targeted \clustername, resulting in spectroscopic redshifts for the cluster and eight cluster member galaxies (Section~\ref{sec:cluster}). Slits were placed on at least one image of each of the lensed sources. These spectra resulted in a redshift limit of $1.78<z<3.9$ based on weak continuum and lack of spectroscopic features in the spectra within the wavelength coverage of the data, $\Delta\lambda = 5920-10350$~\AA. We note that \citet{Bayliss2016} provide a ``best guess'' redshift for image 2.2 (A.1 in their notation) $z=2.1955$, based on very weak spectroscopic features in emission (their Figure~9, panel c), but caution that these may be misidentified. The FIRE data, described below, rule out this solution. 

\paragraph{Magellan/LDSS3} We obtained multislit spectroscopy of the lensed images using the Magellan Clay telescope with the Low Dispersion Survey Spectrograph (LDSS3-C) on 2018 Jan 9 (PI: Sharon).
The observations were conducted under good conditions with sub-arcsecond seeing, and slits were placed on eight out of the nine lensed images. However, none of the observations resulted in a redshift measurement for the multiply-imaged systems, due to the wavelength coverage of the instrument and the absence of strong enough \Lya\ emission. 
\paragraph{Magellan/FIRE} Near-IR spectroscopy yielded robust spectroscopic redshifts for several objects of interest in the field.
We observed multiple sources in the core of \clustername\ with the Folded-port Infrared Echellette (FIRE; \citealt{Simcoe2013}) spectrograph at the Magellan-I Baade telescope.
Observations took place on 2018 Jan 28-29 (PI: Gladders)
; the median seeing during the time of observation was $1\farcs0$, and the airmass ranged between 1.1--2.0.

In total, we observed five different positions in the field using the $1\farcs0 \times 6\farcs0$ slit, with FIRE in high resolution echelette mode. The slit was set at position angles chosen to allow a clean nod of $2\farcs0$ along the slit between neighboring science exposures, and two of the slit positions yielded traces from two sources of interest. With the $1\farcs0$ wide slit FIRE delivers spectra with a resolution of $R = 3600$ ($\sigma_v$ $=83$  \kms) and covers a wavelength range of 0.82--2.5 $\mu$m in a single-object cross-dispersed setup \citep{Simcoe2008}. We reduced the data using the FIRE reduction pipeline (FIREHOSE)\footnote{\url{http://web.mit.edu/~rsimcoe/www/FIRE/ob_data.htm}}; our observations resulted in clear astrophysical emission lines in sky-subtracted 2D spectra for seven distinct sources observed across the five different slit positions, and no continuum emission detection. For emission line sources FIREHOSE allows manual identification of source traces using individual emission lines. The user-supplied line positions and trace location are combined with a trace model to extract object spectra by jointly fitting the source trace along with the 2-dimensional sky spectrum using the source-free regions along the slit. We also performed observations of A0V telluric standard stars during the night of our science observations and at similar airmass. The A0V spectra were used to calibrate the extracted science spectra \citep{Vacca2003} using the xtellcor procedure as a part of the spextool pipeline \citep{Cushing2004}, which is called as a part of the FIREHOSE reduction process. 

We measured cosmological redshifts for each source with an extracted FIRE spectrum by identifying families of nebular emission lines---H$\alpha$, H$\beta$, H$\gamma$, [N II] at $\lambda\lambda$6855, [O III] at $\lambda\lambda$4960,5008, and [O II] at $\lambda\lambda$3727,3729---and fitting a Gaussian profile to each emission line. We estimated the mean redshift for each spectrum as the average of the individual line redshifts, and the uncertainty as the quadrature sum of the uncertainties in the individual line centroids from each Gaussian profile, the uncertainty in the wavelength solution (always highly sub-dominant), and the scatter in the measured redshifts of the individual emission lines. Individual source redshifts are labeled in Figure~\ref{fig:cluster} with solid circles, given in Table~\ref{tab:FIREspeclines} and the extracted emission line spectra of those sources are shown in the appendix Figures~\ref{fig:jellyfish}, \ref{fig:FIREsys1}, \ref{fig:FIREsys2} \ref{fig:FIREsys3}.

\startlongtable
\begin{deluxetable}{c|cc}

\tablecaption{Emission lines detected in the FIRE spectra for the lensed systems as\label{tab:FIREspeclines}}
\tablehead{
\colhead{system} & \colhead{z} & \colhead{Restframe emission lines} 
}
\startdata
system 1  & 2.363 &  [O III]\OIII \\
\hline
system 2 & 2.364 &[O II]\OII\ \\
         &  & [O III]\OIII\ \\
         &  & \Halpha 6563\\
\hline
system 3 & 3.048  & [O III]\OIII\  \\
       &  & \Hgamma 4340  \\
\enddata
\end{deluxetable}



\section{Selection of cluster members}
\label{sec:members}
\begin{figure}
\begin{center}
   \includegraphics[trim=0 0.5cm 0 0,clip,width=0.45\textwidth]{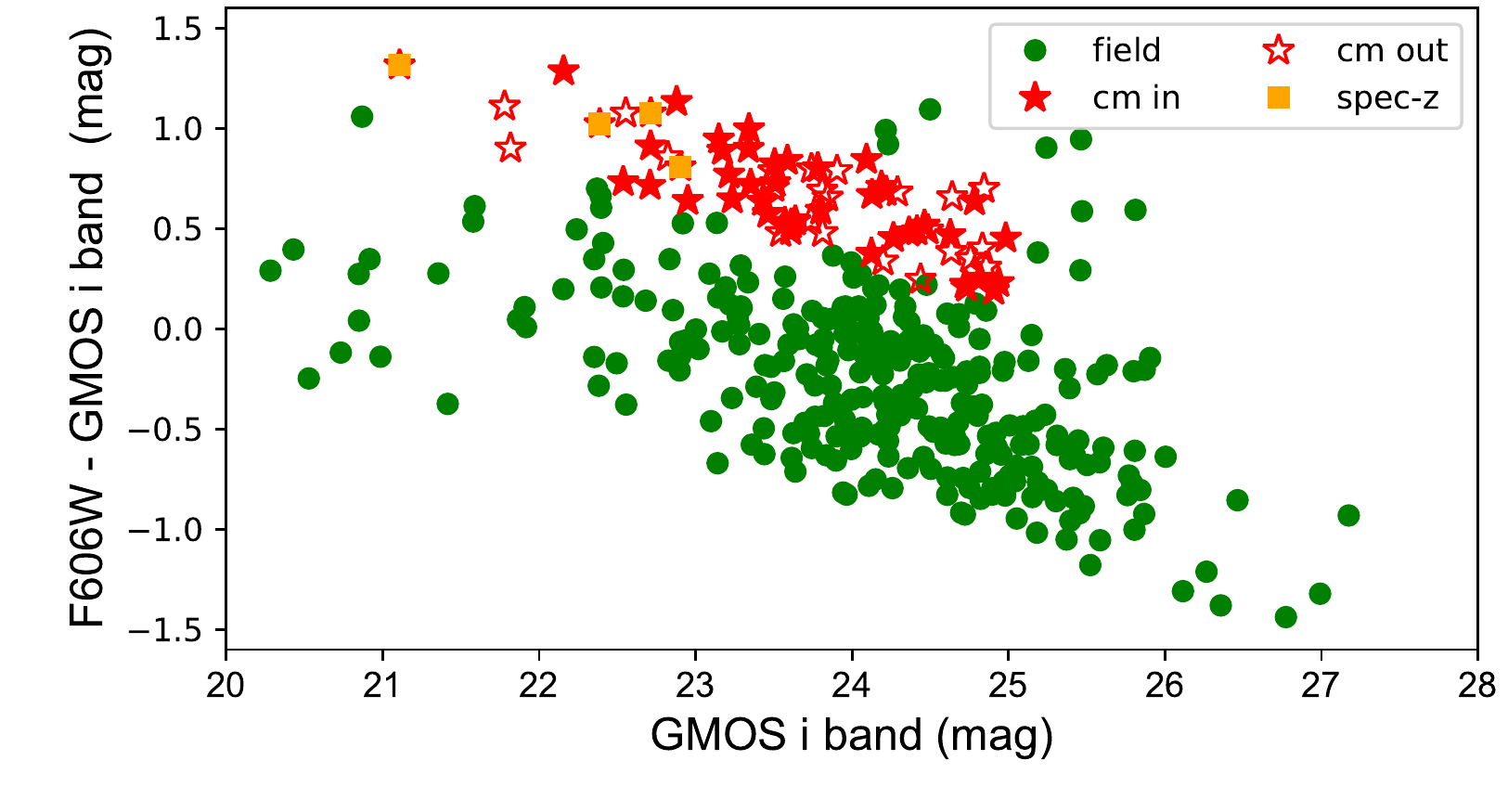}
\end{center}
        \caption{Color-magnitude diagram of galaxies within the ACS field of view. The galaxies that are selected as the red sequence are marked with red star symbols; filled symbols are galaxies within \CMR\ from the BCG. Galaxies with spectroscopic redshifts at the cluster redshift are labeled in yellow.}
    \label{fig:redsequence}
\end{figure}

We identified cluster-member galaxies by color, using the red sequence technique \citep{Gladders2000} from a GMOS $i$-band and \HST/ACS F606W color-magnitude diagram (Figure~\ref{fig:redsequence}). To measure galaxy colors, we first aligned and re-sampled the GMOS $i$-band image to match the ACS pixel frame, and used Source Extractor \citep{SEx} in dual-image mode with the GMOS-$i$-band as the detection image. 
Magnitudes were measured as MAG\_AUTO\footnote{\url{https://www.astromatic.net/pubsvn/software/sextractor/trunk/doc/sextractor.pdf}} within the $i$-band detection aperture in both images. 

Stars and other artifacts were rejected from the catalog based on their location in a MU\_MAX vs MAG\_AUTO diagram in the \hst\ photometry.
The BCG and the other spectroscopically-confirmed galaxies \citep{Bayliss2016} were used to identify the red sequence locus in color-magnitude space. 
We include in the cluster-member catalog galaxies brighter than $i=25$ mag that lie within \CMR\ in projection from the BCG, where both the ACS and GMOS images have complete coverage. Being far from the center, the galaxies in the outskirts do not have a significant impact on the mass at the cluster core or the lensing configuration. Figure \ref{fig:redsequence} shows the selection made for cluster member galaxies.

We attempted constructing cluster-member catalogs with and without convolving the \hst\ image with the much larger GMOS point spread function (psf). We find that while the red sequence becomes more diffuse due to contamination from bright nearby objects, the overall selection of cluster members is not significantly affected. After examining the discrepancies, we conservatively choose to use the photometry based on the natural resolution of the \hst\ image to reduce contamination. We note that only two faint galaxies (with $i>24$) near the cluster core are marginally near the color-magnitude cut, and would be selected by the psf-matched procedure. High-resolution near-IR data would be required to unambiguously determine cluster membership. Our final cluster member catalog contains 45 
galaxies within \CMR\ of the BCG. 
The selected cluster members galaxies are marked with yellow ellipses in Fig.~\ref{fig:PISCO}.

\section{Lensing Analysis and Mass Models} 
\label{sec:lensing}

\subsection{Lens Modeling Methodology}
\label{Methodology}
We compute a mass model of the core of \clustername\ from the strong lensing evidence, using the publicly available lensing algorithm \lenstool\ \citep{Jullo2007}. 
We refer the reader to \cite{Kneib1996}, \cite{Smith2005}, \cite{Verdugo2011} and \cite{Richard2011} for more details on the strong lens modeling approach used in this work. This section provides a short summary.  
We model the cluster mass distribution as a series of dual pseudo-isothermal ellipsoid (dPIE, \citealt{Eliasdottir2007}) parametric mass halos, with seven free parameters: the position $\Delta\alpha$, $\Delta\delta$; ellipticity $\epsilon$; position angle $\theta$; normalization $\sigma_0$; truncation radius $r_{cut}$; and core radius $r_{core}$. 
We use as constraints the positions of prominent emission clumps in each lensed image, and the spectroscopic redshifts of the lensed sources (see Section~\ref{sec:cstr}). The \lenstool\ algorithm uses a Monte Carlo Markov Chain (MCMC) formalism to explore the parameter space. It identifies the best fit as the set of parameters that minimize the scatter between the observed and predicted image-plane positions of the lensed features. 

The lens plane is modeled as a combination of cluster-scale and galaxy-scale dPIE halos. For the cluster-scale DM halos, we fix the truncation radius ($r_{cut}$) at 1500 kpc, since it is too far beyond the strong lensing regime to be constrained by the strong lensing evidence. The other parameters are generally allowed to be solved for by the lens model, unless otherwise indicated. 

Galaxy-scale halos represent the contribution to the lensing potential from cluster member galaxies (Section~\ref{fig:redsequence}). Their positional parameters ($\Delta\alpha$, $\Delta\delta$; $\epsilon$; $\theta$) are fixed to their observed values as measured with Source Extractor \citep{SEx}. To keep the number of model parameters manageable, the slope parameters of the galaxy-scale potentials are scaled to their observed $i$-band luminosity with respect to L$^*$, using a parametrized mass-luminosity scaling relation (see \citealt{Limousin2007} and discussion therein on the validity of such parametrization) leaving only $r_{cut}$ and the central velocity dispersion ($\sigma_{0}$) free to vary. 
The BCG is modeled separately, since we do not expect it to necessarily follow the same scaling relation (\citealt{Newman2,Newman1}). 

Although likely a cluster member, a bright star-forming ``jellyfish'' galaxy \citep{Ebeling2014} that appears $7\farcs0$ east of the BCG
is not included in the cluster-member catalog, as its brightness significantly deviates from the mass-luminosity relation of the passive cluster member galaxies. 
This galaxy is far enough from the multiply-imaged systems to not significantly affect the lensing configuration, and for the purpose of the lensing analysis it mainly contributes a small increase in the total mass, which is expected to  be degenerate with the cluster scale DM clump. We discuss this galaxy in Appendix~\ref{sec:jellyfish}.

As explained below in Section~\ref{sec:dm}, we consider several lens models, each with a different number of cluster-scale DM halos and modeling assumptions. In all cases, we require that the number of free parameters is smaller than or equal to the number of constraints. We note however that given the small number of lensed sources observed with the existing data, the model may be under-constrained even if this criterion is formally satisfied. 

The lensing constraints (Section~\ref{sec:cstr}) come from the identified image-plane locations of multiple images of the lensed sources, and individual emission knots within each galaxy. The multiple constraints within each galaxy assist in constraining the lensing parity, and provide leverage over the relative magnification between the images, without explicitly using the flux ratios as constraints. The latter can sometimes be affected by variability or microlensing \citep[e.g.,][]{Fohlmeister2008,Dahle2015}.

\subsection{Lensing constraints}
\label{sec:cstr}
 \citet{Bayliss2016} identified three arc candidates in the field of \clustername, and constrained their redshift to the range $1.78 < z < 3.9$ based on lack of emission lines in their Gemini-GMOS spectroscopy.
Here, we refine the identification and report nine lensed images of three distinct sources in the field of \clustername. Each source has three lensed images. As described in Section~\ref{sec:spectroscopy}, we obtained spectroscopic redshifts of at least one image in each system. Constraining the model with spectroscopic redshifts is crucial for a precise and accurate lens model \citep{Johnson2016}. 

While ground-based data can reveal strong-lensing evidence, accurate modeling of the mass distribution requires \HST\ resolution to precisely select multiply-imaged features used as constraints. The positions of the images are marked in Figure~\ref{fig:cluster}, color-coded by system; the inset shows a zoomed-in view of the three systems. The \hst\ data resolve the galaxies and reveal their internal morphology; system 2 and system 3 show clear distinct emission knots that we use as constraints in order to better probe the mass profile of the galaxy cluster. Table~\ref{tab:cstr} summarizes the positions and the spectroscopic redshifts of these systems.
The unique morphology of system~1 and system~2, and the identical morphology observed in their multiple images, result in a robust identification even without spectroscopic confirmation of all three images of each system as was obtained for system~3.

In addition to the secure, spectroscopically confirmed multiply-imaged galaxies, we identify three candidate  multiply-imaged systems. Since the strong lensing model is used to help identify these candidates, we do not use those systems as constraints. Table~\ref{tab:candidate} indicates the position of the identified multiple image candidate systems.

\startlongtable
\begin{deluxetable}{c|ccccc}
\tablecaption{Secure multiply-imaged systems. System refers to the name of the group of images coming from the same source galaxy. ID refers to the name of the image. R.A. and Decl. are the right ascension and declination position (J2000) of the image. Spec-$z$ is the measured spectroscopic redshifts. Rms refers to the distance in the image plane between the projected geometrical center of the source and the image.\label{tab:cstr}
}
\tablehead{
\colhead{system} & \colhead{ID} & \colhead{R.A.} & \colhead{Decl.} & \colhead{{\it spec}-z} & \colhead{rms} \\
\colhead{ } & \colhead{ } & \colhead{hh:mm:ss} & \colhead{dd:mm:ss} & \colhead{ } & \colhead{''}
}

\startdata
system 1 & 1.1 & 3:56:20.458 & -53:37:53.265 & \zspecA\ & 0.11 \\
& 1.2 & 3:56:20.484 & -53:37:50.799 & \nodata & 0.23 \\
 & 1.3 & 3:56:21.317 & -53:37:38.073 & \nodata & 0.27 \\
\hline
system 2 & 2.1 & 3:56:20.239 & -53:37:53.951 & \zspecB\ & 0.05 \\
& 2.2  & 3:56:20.336 & -53:37:47.965 & \nodata  & 0.04 \\
& 2.3  & 3:56:20.952 & -53:37:38.157 & \zspecB\ & 0.01 \\
& 21.1 & 3:56:20.235 & -53:37:53.329 & \zspecB\ & 0.06 \\
& 21.2 & 3:56:20.302 & -53:37:48.898 & \nodata  & 0.12 \\
& 21.3 & 3:56:20.945 & -53:37:38.001 & \zspecB\ & 0.04 \\
& 22.1 & 3:56:20.230 & -53:37:53.101 & \zspecB\ & 0.12 \\
& 22.2 & 3:56:20.291 & -53:37:49.157 & \nodata  & 0.08 \\
& 22.3 & 3:56:20.937 & -53:37:37.886 & \zspecB\ & 0.04 \\
\hline
system 3 & 3.1 & 3:56:19.895 & -53:37:59.115 & \zspecC\  & 0.05 \\
& 3.2  & 3:56:20.123 & -53:37:44.328 & \zspecC\ & 0.19 \\
& 3.3  & 3:56:20.562 & -53:37:37.430 & \zspecC\ & 0.19 \\
& 31.1 & 3:56:19.871 & -53:37:58.899 & \zspecC\ & 0.05 \\
& 31.2 & 3:56:20.098 & -53:37:44.510 & \zspecC\ & 0.19 \\
& 31.3 & 3:56:20.532 & -53:37:37.370 & \zspecC\ & 0.20 \\
& 32.1 & 3:56:19.868 & -53:37:58.629 & \zspecC\ & 0.00 \\
& 32.2 & 3:56:20.085 & -53:37:44.730 & \zspecC\ & 0.27 \\
& 32.3 & 3:56:20.540 & -53:37:37.130 & \zspecC\ & 0.19 \\
\enddata

\end{deluxetable}

\startlongtable
\begin{deluxetable}{c|ccc}

\tablecaption{Candidate multiply-imaged systems. System refers to the name of the group of images coming from the same source galaxy. ID refers to the name of the image. R.A. and Decl. are the right ascension and declination position (J2000) of the image. \label{tab:candidate}}
\tablehead{
\colhead{system} & \colhead{ID} & \colhead{R.A.} & \colhead{Decl.} \\
}

\startdata
system 4  & 4.1 & 3:56:19.093 & -53:37:56.703 \\
& 4.2 & 03:56:18.403 & -53:37:51.059  \\
& 4.3 & 3:56:20.222 & -53:37:29.576  \\
\hline
system 5 & 5.1 & 3:56:22.947 & -53:37:53.300 \\
& 5.2 & 3:56:22.864 & -53:37:57.408 \\
& 5.3 & 03:56:21.842 & -53:38:06.655 \\ 
\hline
system 6 & 6.1 & 3:56:24.330 & -53:38:10.668 \\
& 6.2 & 3:56:24.367 & -53:38:10.358 \\
\enddata
\end{deluxetable}

\subsection{Dark Matter Halos}
\label{sec:dm}
We compute four models with one or two DM halos and varying free parameters, to investigate the spatial distribution of DM in the cluster core with respect to the stellar component. These models are summarized in Table~\ref{tab:fitstat}, and described below.

The spatial distribution of cluster-member galaxies appears to be separate in two components, with a concentration of cluster-member galaxies grouped $\sim$ 150 kpc West of the BCG. The formation of the arcs between the BCG and this concentration of galaxies indicates that the underlying dark matter mass distribution of the cluster may also shows a two-components structure. Similar lensing configurations are seen in several lower-redshift clusters, whose lens models are dominated by two cluster-scale DM halos \citep[e.g.,][]{sharon2019}. To test the hypothesis that this cluster is also dominated by two halos, we compute two sets of lens models: 
The first set of models, labeled A and B, have one cluster-scale DM halo (DM1 in Table~\ref{tab:fitstat}). A second set of models, labeled C and D, have two cluster-scale DM halos (DM1 and DM2 in Table~\ref{tab:fitstat}). In models A and C, the center of DM1 is not fixed, adding two free parameters to these models. Contribution of cluster-member galaxies is included in all models in the same way, as explained in \ref{Methodology}.
DM1 is assumed to be located at or near the position of the BCG, for two reasons; First, the BCG presents a regular luminosity profile which suggests that it is not disturbed and therefore this galaxy would be at the cluster center. Second, we lack the ability to properly constrain the Eastern extent of the cluster since we do not identify secure lensing constraints in this region at the depth of the existing data.
The position of DM2 
is free with a loose prior that positions it around the group of galaxies on the western part of the cluster. We chose that location as it is more likely that the DM clump is located close to a luminous counterpart \citep{Broadhurst2000,Broadhurst2005}.

The second hypothesis we investigate is whether the BCG and its associated DM halo sit at the center of the cluster-scale DM halo. 
To test this scenario, we fix the position of DM1 to the position of the BCG in models B and D.


\subsection{Lens modeling results}
Table~\ref{tab:fitstat} lists the best-fit values of the lens model parameters for each one of the four test models.
To evaluate the lens models, we employ two statistical criteria, as described below. One criterion is named rms and represents the average difference between the observed position of a multiple image and the predicted position from the geometrical center of the best-fit model in
the image plane given in arcseconds. Thus we seek to reduce the rms as much as possible.
The models with one cluster-scale DM clump show an rms of $0\farcs3$ and $0\farcs1$ for the fixed and free dark matter halo respectively. 
The models with two cluster-scale DM clumps result in better rms of $0\farcs06$ and $0\farcs07$ for the fixed and free dark matter halo, respectively. 

The rms criterion suggests that the models with two cluster-scale DM halos are significantly better; however, this criterion does not account for the increased flexibility due to the additional free parameters. To account for that,
we further evaluate the models using a second criterion, the Bayesian Information Criterion (BIC), which was presented in previous works \citep[see section 5.1 in][]{Mahler2018, Lagattuta2017, Acebron2017}. 
The BIC enables a quantitative comparison between similar models; it is a statistical measurement based on the model likelihood $\mathcal{L}$, penalized by the number of free parameters $k$ and the sample size $n$ (i.e. $2 \times$ the number of multiple images): 
\begin{equation}
{\rm BIC} = -2 \times \log(\mathcal{L}) + k \times \log(n),
\end{equation}
We seek to maximize the likelihood (or reduce the first term of Equation 1). However, arbitrarily increasing the number of free parameters would overfit the data.
The second term provides means of balancing the over-fitting. It represents a combination of the number of constraints and the number of free parameters and increases the global value of the BIC. We seek the lowest BIC possible. Using the BIC will help us estimate the improvement of the likelihood in comparison with the freedom allowed by the new parameters such as the secondary halo or freeing the position of
DM1. 
The number of constraints is identical among all models. It is recommended to only compare similar models because otherwise the likelihood will not be a similar description of the model performance. 

We discuss here the performance of the four models. To avoid a possible confusion we will refer to the model letter as listed in Table~\ref{tab:fitstat}.

Model A, with one cluster halo at a fixed position, has an rms of $0\farcs3$ and a BIC of 20. Freeing the halo center as was done in model B reduces the rms to $0\farcs1$ and the BIC to $-$16. This is a considerable improvement, indicating that according to both the rms and the BIC criteria model B, with a free halo, is a better fit to the lensing constraints than model A.

We compare models A and C --- the two models with fixed ``main'' DM halo, but with and without a secondary halo around the group, respectively. The two-halo model results in a drastically reduced rms of $0\farcs06$ and a lower BIC of -4. However, we caution that such a low rms is unrealistic in comparison to other well-constrained models in the literature, and may suggest overfitting. 
It is possible that model C is too flexible, and the lack of constraints west of the group allows it to 
compensate for the fixed position of DM1 with DM2. We note that its mass distribution
is drastically different from the three other models as shown in
Figure\,\ref{fig:masscontours} especially in the regions that lack lensing constraints. We conclude that we currently do not have enough information to properly constrain the position of DM2.

We compare models B and D --- both with free ``main'' DM halo, but with and without a secondary halo around the group. The mass distributions of the two models are similar, as shown in Figure\,\ref{fig:masscontours}. Model D adds a second DM halo near the group, however, it adds only little mass to the model -- as can be inferred by its low normalization parameter $\sigma_0$ (Table~\ref{tab:fitstat}). 
The rms of model B ($0\farcs1$) is similar to the one of model D ($0\farcs07$) even if both remain low in comparison with well constraints cluster in the literature. However the BIC of model B ($-$16) is significantly better than the one for model D (4). This implies that the modeling flexibility offered by the addition of the secondary halo is not required in order to improve the overall goodness of the model.

Models C and D have a similar rms values. Model D has a slightly higher BIC, indicating that there is only little statistical difference between models C and D.
Moreover, as can be seen in Table~\ref{tab:fitstat}, a fixed versus free position of DM1 results in significantly different positions for DM2, due to the location of the lensing constraints between these two halos. As noted above, with the current lensing evidence, the position of DM2 is severely under-constrained. Further lensing observables west of the group are needed in order to constrain the position of DM2, which will make models using the same assumptions as C and D more reliable.

In conclusion, given the available constraints, the BIC criterion identifies model B as the one that compromises best between the goodness of the fit and number of constraints and free parameters. 
Nevertheless, models with two cluster-scale DM halos are not ruled out.

While our statistical assessment suggests that there could be more than one unique ``best'' model that satisfies the lensing constraints, our main conclusions are not significantly affected by the choice of model: The image configurations, regardless of the modeling choices, require that that there be two main mass clumps. None of the modeling results differ on that. We discuss this further in section~\ref{sec:discussion}.

\begin{table*}[h]
\centering
\caption{Candidate Lens Models and Best-Fit Parameters}
\label{tab:fitstat}
\begin{tabular}{lrcccccccc}
\hline 
Model name & Component & $\Delta\alpha^{\rm ~a}$ & $\Delta\delta^{\rm ~a}$ & $\varepsilon^{\rm ~b}$ & $\theta$ & $\sigma_0^{\rm ~c}$ & r$_{\rm cut}^{\rm ~c}$ & r$_{\rm core}^{\rm ~c}$\\ 
(Fit statistics) & -- & (\arcsec) & (\arcsec) &   & ($\deg$) & (\kms) & (kpc) & (kpc)\\ 
\hline 
A -- 1 cluster scale halo, fixed & DM1 & [0.0] & [0.0] & 0.82$^{+0.00}_{-0.02}$ & 39.1$^{+2.0}_{-1.3}$ & 623$^{+22}_{-28}$ & [1500.0] & 1.1$^{+1.4}_{-0.0}$\\ 
rms = 0\farcs3  & BCG & [0.0] & [0.0] & [0.27] & [52.1] & 384$^{+90}_{-140}$ & 7$^{+21}_{-2}$ & [0.6]\\ 
 BIC = 20, AICc = 10  & $L^{*}$ Galaxy & -- & -- & -- & -- & 156$^{+5}_{-5}$ & 56$^{+9}_{-8}$ & --\\ 
 $\log$($\mathcal{L}$) = 5, k= 8, n= 42 & -- & -- & -- & -- & -- & --  & --  & --\\
\hline 
\hline 
B -- 1 cluster scale halo, free & DM1 & 2.9$^{+1.7}_{-3.3}$ & 3.8$^{+0.7}_{-1.4}$ & 0.65$^{+0.06}_{-0.11}$ & 26.0$^{+2.7}_{-3.5}$ & 730$^{+85}_{-47}$ & [1500.0] & 9.8$^{+1.6}_{-1.6}$\\ 
rms = 0\farcs1  & BCG & [0.0] & [0.0] & [0.27] & [52.1] & 498.1$^{+1.7}_{-31.0}$ & 24$^{+23}_{-9}$ & [0.6]\\ 
 BIC = -16, AICc = -27 & $L^{*}$ Galaxy & -- & -- & -- & -- & 116$^{+8}_{-15}$ & 76$^{+109}_{-22}$ & --\\
  $\log$($\mathcal{L}$) = 27, k= 10, n= 42 & -- & -- & -- & -- & -- & --  & --  & --\\
\hline 
\hline 
C -- 2 cluster scale halos, fixed & DM1 & [0.0] & [0.0] & 0.69$^{+0.01}_{-0.11}$ & 33.3$^{+3.2}_{-1.4}$ & 801$^{+53}_{-49}$ & [1500.0] & 3.7$^{+0.9}_{-1.4}$\\ 
rms = 0\farcs06  & BCG & [0.0] & [0.0] & [0.27] & [52.1] & 100$^{+80}_{-132}$ & 54$^{+31}_{-4}$ & [0.6]\\ 
 BIC = -4, AICc = -14  & DM2 & 37.1$^{+1.4}_{-6.6}$ & 3.1$^{+1.1}_{-2.0}$ & 0.84$^{+0.03}_{-0.12}$ & 157.3$^{+6.7}_{-13.6}$ & 560$^{+52}_{-58}$ & [1500.0] & 3.8$^{+0.3}_{-3.0}$\\ 
$\log$($\mathcal{L}$)= 28, k= 14, n=42  &$L^{*}$ Galaxy & -- & -- & -- & -- & 164$^{+77}_{-26}$ & 8$^{+48}_{-1}$ & --\\ 
\hline 
\hline 
D -- 2 cluster scale halos, free & DM1 & 1.9$^{+2.0}_{-2.9}$ & 2.2$^{+1.2}_{-1.3}$ & 0.6$^{+0.01}_{-0.21}$ & 27.3$^{+5.1}_{-3.1}$ & 661$^{+88}_{-89}$ & [1500.0] & 5.6$^{+1.8}_{-3.7}$\\ 
rms = 0\farcs07  & BCG & [0.0] & [0.0] & [0.27] & [52.1] & 419$^{+3}_{-54}$ & 47$^{+24}_{-6}$ & [0.6]\\ 
 BIC = 4, AICc = -3 & DM2 & 21.5$^{+5.4}_{-8.6}$ & 8.3$^{+3.7}_{-3.7}$ & 0.87$^{+0.22}_{-0.36}$ & 165.6$^{+21.4}_{-107.4}$ & 232$^{+191}_{-123}$ & [1500.0] & 1.4$^{+1.2}_{-2.0}$\\ 
 $\log$($\mathcal{L}$) = 28, k= 16, n= 42& $L^{*}$ Galaxy & -- & -- & -- & -- & 104$^{+13}_{-16}$ & 106$^{+143}_{-23}$ & --\\ 
\hline 
\end{tabular}
\medskip\\
Quantities in brackets are fixed parameters~~~~~~~~~~~~~~~~~~~~~~~~~~~~~~~~~~~~~~~~~~~~~~~~~~~~~~~~~~~~~~~~~~~~~~~~~~~~~~~~~~~~~~~~~~~~~~~~~~~~~~~~~~~\\[1pt]
We report statistical quantities such as the Bayesian Information criterion (BIC), the corrected Akaike information criterion (AICc), the likelihood $\log$($\mathcal{L}$) the number of free parameter k, and the sample size n. ~~~~~~~~~~~~~~~~~~~~~~~~~~~~~~~~~~~~~~~~~~~~~\\[1pt]

$^{\rm a}$ $\Delta\alpha$ and $\Delta\delta$ are measured relative to the reference coordinate point: ($\alpha$ = 59.0896383, $\delta$ = -53.6310962)~~~~~~~~~~~~~~~~~~~~~~~~~~~~~~\\[1pt]
$^{\rm b}$ Ellipticity ($\varepsilon$) is defined to be $(a^2-b^2) / (a^2+b^2)$, where $a$ and $b$ are the semi-major and semi-minor axes of the ellipse~~~~~~~\\[1pt]
$^{\rm c}$ $\sigma_0$, r$_{\rm cut}$ and r$_{\rm core}$ are respectively the central velocity dispersion, the cut radius, and the core radius as defined for the dPIE potential used in our modelisation. For $L^{*}$ Galaxy this value represent the parameter of the galaxy that we optimised for our mass-to-light ratio. We refer the reader to section \ref{Methodology} for a summary, and \citet{Limousin2007,Eliasdottir2007} for a more detailed description of the potential.~~~~~~~~~~~~~~~~~~~~~~~~~~~~~~~~~~~~~~~~~~~~~~~~~~~~~~~~~~~~~~~~~~~~~~~~~~~~~~~~~~~~~~~~~~~~~~~~~~~~~~~~~~~~\\[1pt]

\end{table*}

\section{discussion}
\label{sec:discussion}

Optical imaging and spectroscopy of \clustername\ indicate that it has a two-components distribution of cluster-member galaxies, with two main stellar components separated by 21 arcsec ($\sim$170kpc). Our strong lensing analysis finds that the distribution of DM is consistent with that of the galaxies. In this section, we compare the stellar and DM distributions and discuss some of their unusual properties. 

The GMOS spectra of the BCG and two cluster-member galaxies from a nearby group are shown in Figure\,\ref{fig:gmos_spectra} (retrieved from \cite{Bayliss2016}), in red, blue and green lines, respectively. The velocity offset between the BCG and these other galaxies is $<300$ \kms. The spectrum of the BCG shows [OII] in emission, which is indicative of star formation; the other two galaxies show little to no [OII] emission. Observing star formation in a BCG is not unusual, and overall, the spectroscopic data are consistent with these galaxies arising from a similar population of galaxies at the same redshift. The spectrum of this galaxy does have a velocity offset between the absorption and emission features ($\sim550$ \kms). This offset could trace an in-falling filament of cooling gas in the cluster core. Velocity difference between galaxies are reported here using the measurement of the absorption features. 

\begin{figure}
    \centering
    \includegraphics[width=0.45\textwidth]{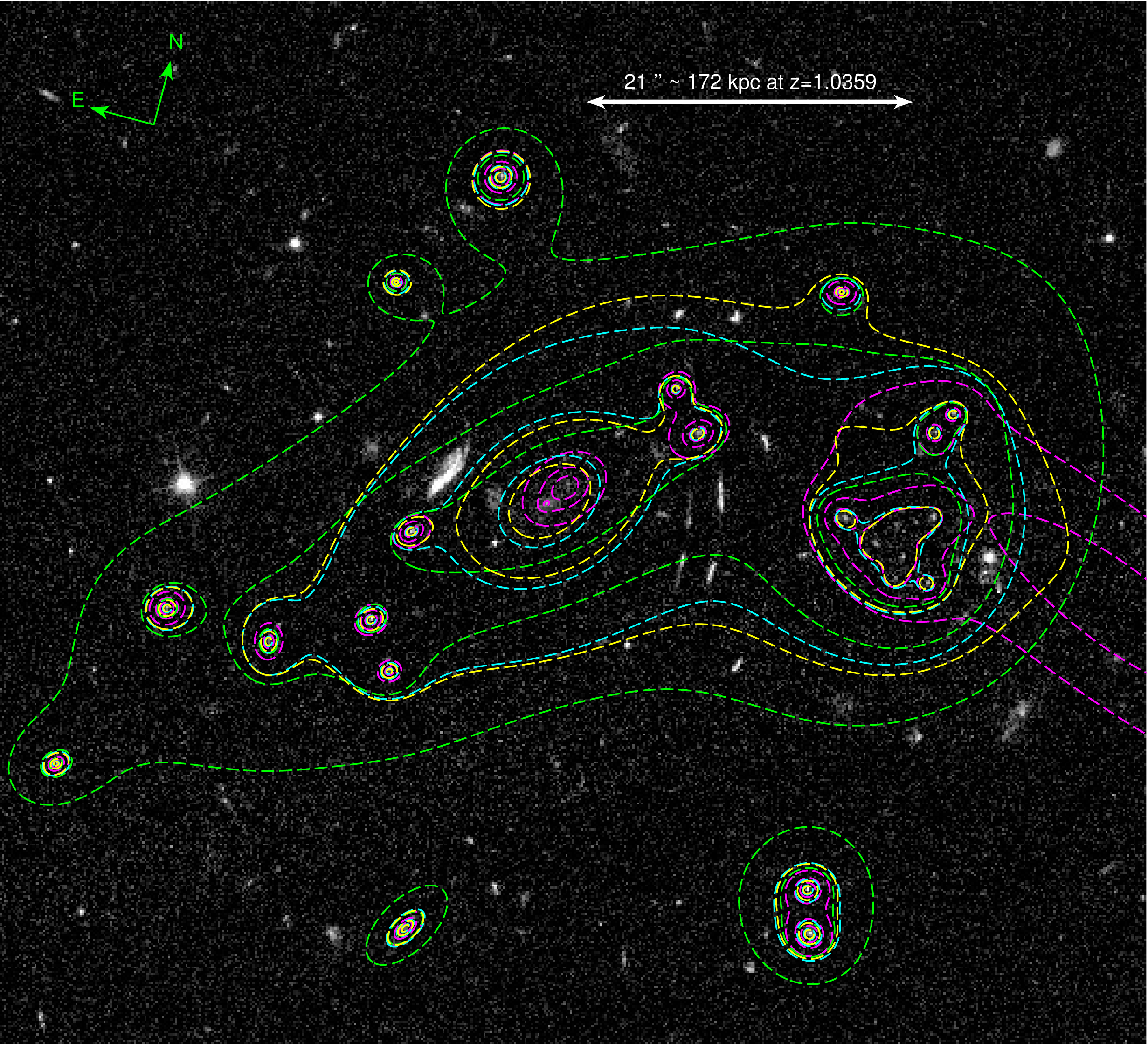}
    \caption{Projected mass density contours from models A (green), B (cyan), C (magenta), and D (yellow). The total projected mass density show two component, regardless of modeling choices, with one clump centered near the BCG and one near the nearby group of cluster-member galaxies. Contours are plotted at 0.5, 1, and 2$\times10^{9}$\msun.kpc$^{-2}$}
    \label{fig:masscontours}
\end{figure}

\begin{figure*}
\begin{center}

   \includegraphics[scale=0.6]{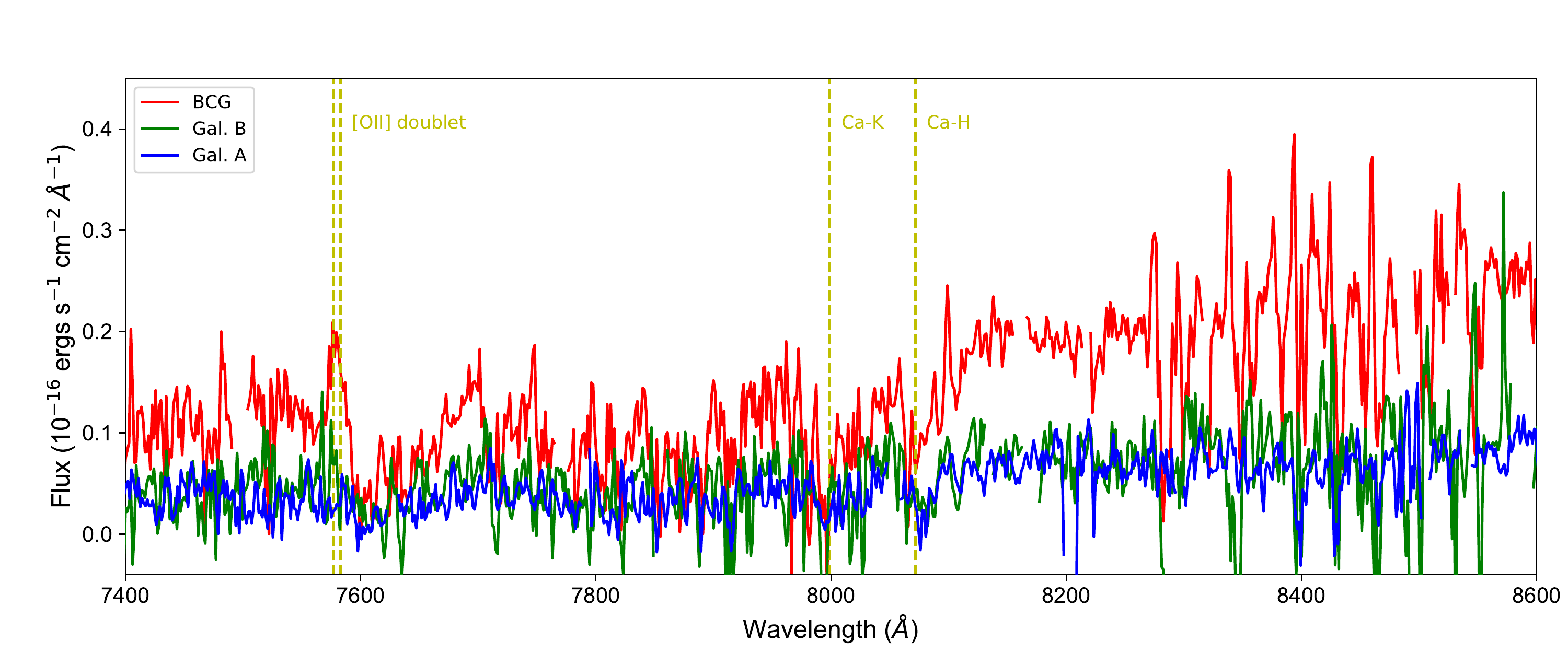}
        \caption{The GMOS spectra of the BCG in red and two other cluster members, part of the group discussed in section~\ref{sec:discussion}, in blue and green. The two galaxies are marked with green squares located in the group of galaxies in Figure~\ref{fig:cluster}. The vertical yellow dashed line marks identified spectral features at the redshifts of the BCG.
        }
    \label{fig:gmos_spectra}
    \end{center}
\end{figure*}

All four strong lensing models are consistent with a two-components mass distribution, with one component centered near the BCG and one near the group. Figure~\ref{fig:masscontours} plots the mass contours derived from model A (green contours), B (cyan contours), C (magenta contours), and D (yellow contours), showing that regardless of modeling choices a secondary mass distribution is needed in order to reproduce the lensing constraints.

We measure the mass of each of the two main structures, by summing the projected mass density within apertures of 80 kpc radius centered on the BCG and on the group. We choose 80 kpc as it 

We report those values in Table \ref{tab:masses}, as well as their ratios, for each of the tested models. We find that regardless of the model used, we always find a similar total projected mass ratio between the two structures.

The small radial velocity offset between the BCG and two measured galaxies in the group (only 300 \kms) strongly suggests that most of the motion is transverse; the separation between the two mass clumps is small, 21 arcsec ($\sim$170 kpc); and their mass ratio is near-equal. Those aspects fulfill most of the criteria laid out by \citet{Dawson2012}, see section \ref{sec:intro}, arguing that this cluster could be a dissociative merger candidate like the Bullet Cluster undergoing a major merger event. Absent additional information on the dynamical state of the hot gas prevent us to be conclusive regarding its state. Indeed, the cluster can either be in a pre-merger state or has gone through a very recent merger. High resolution X-ray data can distinguish between these scenarios. This arrangement, where the BCG is spatially separated from other cluster members is atypical, and supports the conclusion that we are observing a major merger event.

\startlongtable
\begin{deluxetable}{l|cccc}
\tablecaption{The projected masses for our four different models enclosed within a radius of 80 kpc centered on two locations. M$_{\rm BCG}$ refers to the mass measured in a circle center on the BCG. M$_{\rm group}$ refers to the mass measured in a circle center on the group. Ratio refers to the ratio of the two previous column. f$_{CM}$ refers to the fraction of the total mass contain in clusters members, within circle 800kpc. All masses indicated in this table are in units of $10^{12}$\msun.\label{tab:masses}}
\tablehead{
\colhead{Model} & \colhead{M$_{\rm BCG}$ } & \colhead{M$_{\rm group}$ } & \colhead{Ratio} & \colhead{f$_{CM}$}  \\
}
\startdata
A - 1 halo, DM1 fixed  & $18.9_{-0.5}^{+0.0}$ & $15.1_{-0.0}^{+3.3}$ & 1.25 & 0.46 \\
B - 1 halo, DM1 free & $17.0_{-0.3}^{+0.0}$ & $11.0_{-0.0}^{+5.7}$ & 1.55& 0.27 \\
C - 2 halos, DM1 fixed & $15.4^{+1.0}_{-0.0}$ & $10.9^{+5.5}_{-0.0}$ & 1.41& 0.54 \\
D - 2 halos, DM1 free & $16.2 _{-1.1}^{+0.0}$ & $10.2_{-0.0}^{+4.8}$ & 1.58 & 0.28 \\
\enddata

\end{deluxetable}


Examination of the two main mass concentrations reveals
that their galaxy content is significantly different. 
One component is dominated by a BCG, with a half-light radius larger than any other cluster member. The other component is composed of a group of eight cluster members. While it is unlikely that this apparent clustering of eight galaxies within $\sim$9\arcsec ($\sim$70 kpc) is only due to a projection effect, larger spectroscopic coverage could tackle this issue, as would a more refined red sequence based solely on \HST\ data.


\begin{figure*}
\begin{center}
   \includegraphics[scale=0.5]{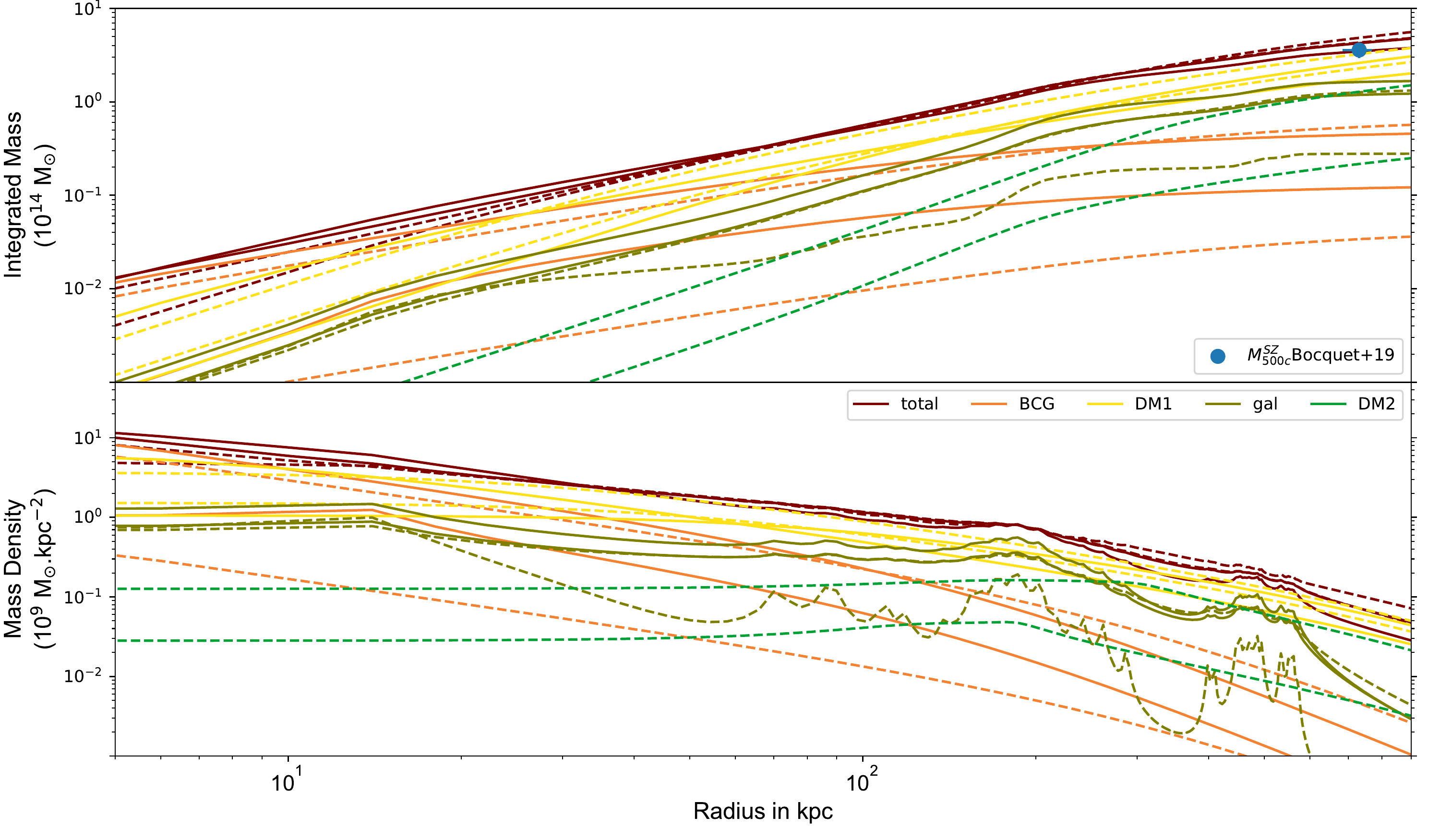}
\end{center}
        \caption{The integrated (upper panel), and differential (lower panel) mass profile of the cluster. The mass profiles are drawn from a circular aperture centered on the BCG. The color coding shows the total mass profiles as well as the different sub-components. The solid and dashed lines represent the set of models with one dark matter halo and two dark matter halos, respectively. The blue dot marks the estimated SZ mass of the cluster from \citet{Bocquet2019}.
        }
    \label{fig:mass_prof}
\end{figure*}

In Figure\,\ref{fig:mass_prof} we present the mass and density profiles for the four different models.  The statistical uncertainties of the models are very low ($<$3\%), and, for clarity, they are not shown in this figure. The error budget is dominated by systematic uncertainties due to various assumptions and modeling choices. 
Our parametric modeling of the cluster mass distribution allows us to isolate the different mass components that contribute to the total mass: 
cluster-scale dark matter clumps, the BCG, 
and the total contribution from the cluster members. We expect degeneracies between the BCG and the core mass of the dark matter halo at least in some of the models, because they are both confined to the same location. The result is a large variation in the mass of the BCG between models, however, this component represents only a small fraction of the total mass. 

The total contribution from cluster member galaxies appears to be a large fraction of the total mass. As can be seen in Table~\ref{tab:masses}, the different models predict that about 27 to 54\% of the total mass is contained in the cluster member galaxies and their associated DM halos. This is resulting from our fitting procedure that allows a constant mass-to-light ratio to vary in amplitude. 
In an analysis of a lower-redshift cluster merger, MACS\,J0417.5$-$154, \citet{Mahler2019} found a significantly smaller ratio between the galaxy and cluster contributions of about 1\%. \citet{Wu2013} investigated this ratio using simulated clusters and reported that up to 20\% of the total mass is contained in subhalos that survived the merger with the main halo, although with large scatter, and strong dependence on formation time. 
In a future analysis of a large sample of clusters, we will investigate whether this is indicative of an evolutionary trend with cosmic time, or anecdotal representation of a larger cluster-to-cluster variation.

Several aspects of this analysis would be better constrained with additional data, primarily multi-band high-resolution imaging, from \hst, and with high resolution X-ray observations. Multi-band \HST\ observations, extending to the near IR, would refine the red-sequence selection of cluster-member galaxies and provide a handle on their stellar mass through spectral energy distribution fitting. These data would also facilitate the detection of new multiple images and the confirmation of image candidates in the east and west parts of the cluster, which are under-constrained with the current data. X-ray observations are necessary for determining the dynamical state of the hot cluster gas. A signature of a shock between the components would indicate a recent major merger \citep{Poole2007}, while X-ray emission from both structures would support a pre-merger scenario.

\section{summary}

We construct a strong lensing mass model of the galaxy cluster \clusterfullname\ at $z=1.0359$, one of the highest-redshift strong lensing clusters known to date. We present spectroscopic confirmation and redshifts of three multiply-imaged lensed galaxies, whose images appear 9.5 to 15 arcseconds west of the BCG. 
The lensed galaxies are spatially resolved, allowing us to use different emission knots in the same system as constraints, which adds leverage on the shape of the mass profile. This provides confidence in our ability to accurately probe the mass distribution, which at the location of the lensing evidence, is constrained to within a few percent.
However, the lack of multiply-imaged systems at the outskirts of the cluster core limits our understanding of the cluster halo.
Nevertheless, \clustername\ appears to be the best-constrained lensing cluster at this redshift bin to date. Other cluster-scale lenses either have too few lensing constraints, are not spectroscopically confirmed, or their apparent lensing evidence is dominated by single galaxies rather than the cluster potential (see section~\ref{sec:intro}).

We employ statistical criteria to evaluate four possible lens models, which are based on different modeling assumptions. 
The lens model indicates that \clustername\ has an Einstein radius of \Erad measured based on the tangential critical curve for a source at $z=3$.
At a radius of 500 kpc the enclosed mass is measured to be \Mfivekpc. We report within a radius of 730 kpc ($\simeq R_{500}$) a value of M$=4.1\pm 0.8\times10^{14} $\msun\,
consistent with the SPT measurements 
\MBocquetfixed\ from \citet{Bocquet2019}. All values are derived from the statistically-favored model (B).

Regardless of modeling assumptions, we find that the projected mass density of the cluster is best described by a two-component mass distribution, with one mass substructure centered around the BCG, and a second mass substructure centered on the observed position of a small group of eight cluster-member galaxies, grouped within a radius of  $\sim$9\arcsec\ ($\sim$70 kpc) diameter circle, located $\sim$170 kpc west of the BCG. The lensing analysis points to a nearly equal mass between the two substructures of \clustername. Nevertheless, the galaxy distribution is significantly different between those mass components -- one dominated by a single galaxy, the other hosting a group of eight galaxies.

The similar masses, the low radial velocity offset between the group and the BCG, and the small impact parameter between the two structures, suggest that this cluster is undergoing a major-merger event. However, to fully characterize this system as a dissociative merger, we would require deep X-ray imaging to probe its intracluster medium and constrain the dynamical state of the cluster gas. If confirmed, \clustername\ will be an important $z>1$ target for next-generation X-ray telescopes.

We find a high mass ratio between the mass associated with cluster-member galaxies to the cluster-scale DM halos at the core of the cluster compared to low mass clusters, perhaps indicating that the subhalos are yet to lose a significant fraction of their DM to the cluster potential. All evidence in hand suggests that \clustername\ provides a unique opportunity to probe the population of high-redshift clusters, and to study the evolution of massive clusters.

\section*{Acknowledgments}

This work is based on observations made with the NASA/ESA {\it Hubble Space Telescope}, using imaging data from GO program 13412.
STScI is operated by the Association of Universities for Research in Astronomy, Inc., under NASA contract NAS5-26555.
This paper includes data gathered with the 6.5 meter Magellan Telescopes located at Las Campanas Observatory, Chile.
Based on observations obtained at the Gemini Observatory, which is operated by the 
Association of Universities for Research in Astronomy, Inc., under a cooperative agreement 
with the NSF on behalf of the Gemini partnership: the National Science Foundation 
(United States), the National Research Council (Canada), CONICYT (Chile), the Australian 
Research Council (Australia), Minist\'{e}rio da Ci\^{e}ncia, Tecnologia e Inova\c{c}\~{a}o 
(Brazil) and Ministerio de Ciencia, Tecnolog\'{i}a e Innovaci\'{o}n
Productiva (Argentina).
TS and JLvdB acknowledge support from the German Federal Ministry of Economics and Technology (BMWi) provided through DLR under project 50 OR 1407. Work at Argonne National  Lab is supported by UChicago Argonne LLC,Operator of Argonne National Laboratory. Argonne, a U.S. Department of Energy Office of Science Laboratory, is operated under contract no. DE-AC02-06CH11357. This research used resources of the Argonne Leadership Computing Facility, which is a DOE Office of Science User Facility supported under Contract DE-AC02-06CH11357.
Work at Argonne National Lab is supported by UChicago Argonne LLC,Operator  of  Argonne  National  Laboratory  (Argonne). Argonne, a U.S. Department of Energy Office of Science Laboratory,  is  operated  under  contract  no.   DE-AC02-06CH11357.

%

 \vspace{5mm}
 \facilities{HST(ACS), SPT, Magellan(LDSS3C, IMACS, PISCO, FIRE), Gemini(GMOS) }



\bibliography{biblio_SPT0356}



\appendix

\section{The jellyfish galaxy}\label{sec:jellyfish}
We identify a galaxy at the cluster redshift ($z=1.017$) that exhibits significant star formation based on its colors, and asymmetric morphology with trails of star formation knots ($\alpha=$3:56:22.310, $\delta=-$53:37:52.700; marked with orange circle in Figure~\ref{fig:cluster}). Galaxies with such properties are often referred to in the literature as ``jellyfish'' galaxies \citep[e.g.,][]{Suyu2010,Ebeling2014}, and are believed to be undergoing stripping as they fall into the intracluster medium, inducing star formation. 
The projected distance between this galaxy and the BCG is $\sim$60~kpc. The redshift corresponds to a velocity offset of $2359$ \kms between the ``jellyfish'' galaxy and the BCG, or 2783 \kms relative to the median cluster redshift $z=$\clusterz. 
As the ratio between the velocity and the velocity dispersion is $v/\sigma_v<2$, and given its small projected distance from the cluster core, it is likely that this galaxy is gravitationally bound to the cluster \citep{Bayliss2017}. 

\begin{figure}[htbp]
\begin{center}
   \includegraphics[scale=0.5]{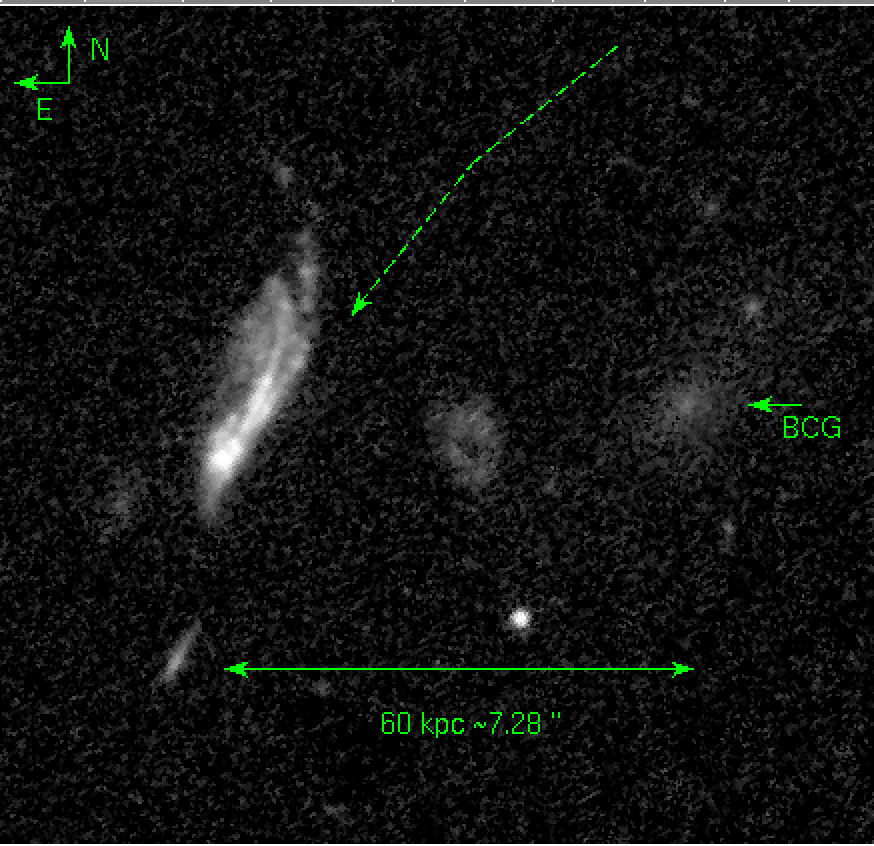}
   \includegraphics[scale=0.5]{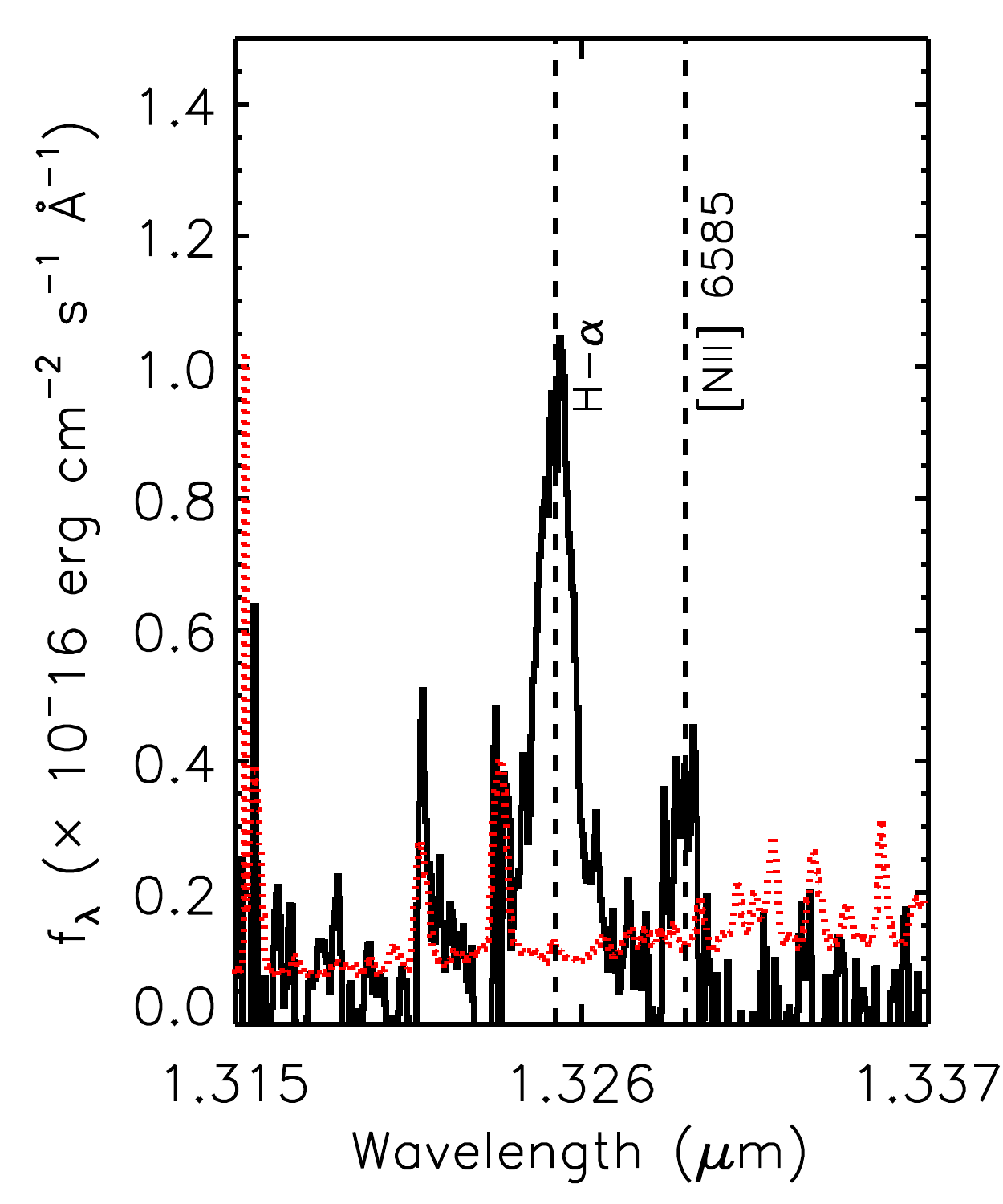}
\end{center}
        \caption{{\it Left:} \HST\ image of the observed jellyfish galaxy in the galaxy cluster, the dashed arrow line represents a projection of the estimated trajectory of the jellyfish galaxy. {\it Right:} Spectral features identified in the FIRE spectra within the slit targeting the jellyfish galaxy.
        }
    \label{fig:jellyfish}
\end{figure}

The apparent stripped gas of the jellyfish suggests that the trajectory on the plane of sky may be from a North-West position. The green dashed arrow shown in figure \ref{fig:jellyfish} represents a best guess of the jellyfish trajectory, following \cite{McPartland2016}. 
The bluer redshift of the jellyfish ($z=1.017$) compared to the cluster ($z=\clusterz$) indicates that the jellyfish galaxy is moving toward us in comparison to the rest of the cluster.

\newpage
\section{FIRE spectra}\label{sec:FIREspectra}
In this Appendix we present sections of the spectra of the lensed galaxies, highlighting the spectral features that were used to measure the spectroscopic redshifts of these galaxies.  
\begin{figure*}[h]
\begin{center}
   \includegraphics[width=0.35\textwidth]{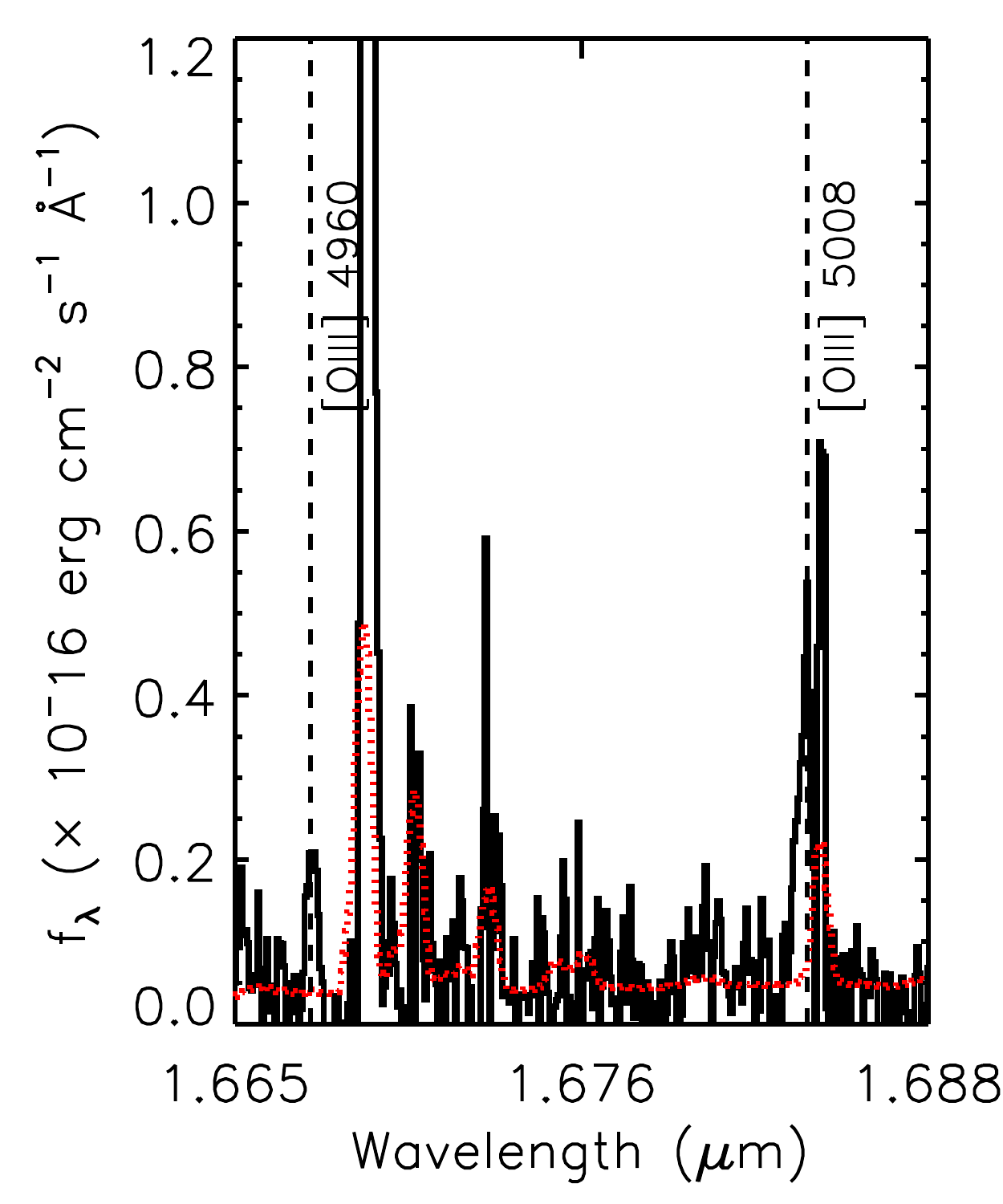}
\end{center}
        \caption{Spectral features identified in the FIRE spectra of image~1.1 at $z=$ \zspecA. }
    \label{fig:FIREsys1}
\end{figure*}

\begin{figure*}
\begin{center}
   \includegraphics[width=0.95\textwidth]{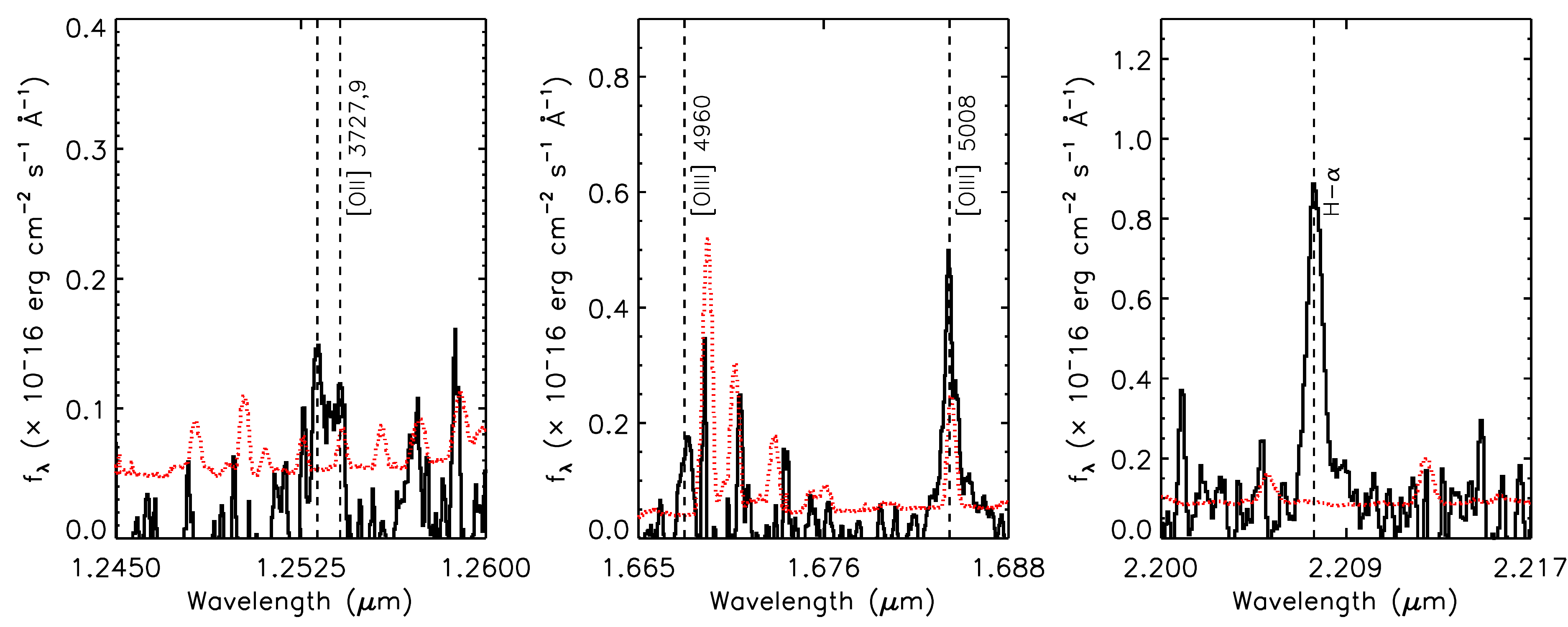}
   \includegraphics[width=0.95\textwidth]{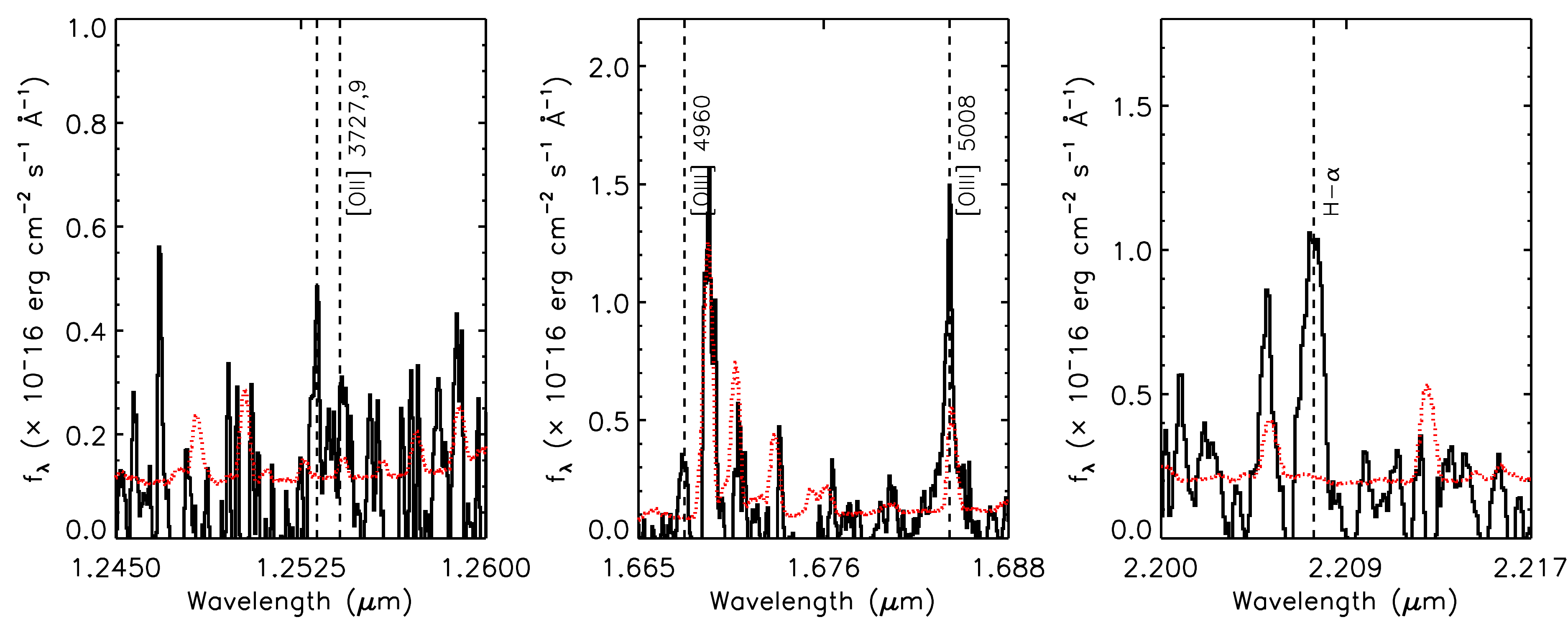}
\end{center}
        \caption{Spectral features identified in the FIRE spectra  for two images of system~2  at $z=$ \zspecB. {\it Top:} image~2.1 - {\it bottom:} image~2.3.}
    \label{fig:FIREsys2}
\end{figure*}

\begin{figure*}
\begin{center}
   \includegraphics[width=0.55\textwidth]{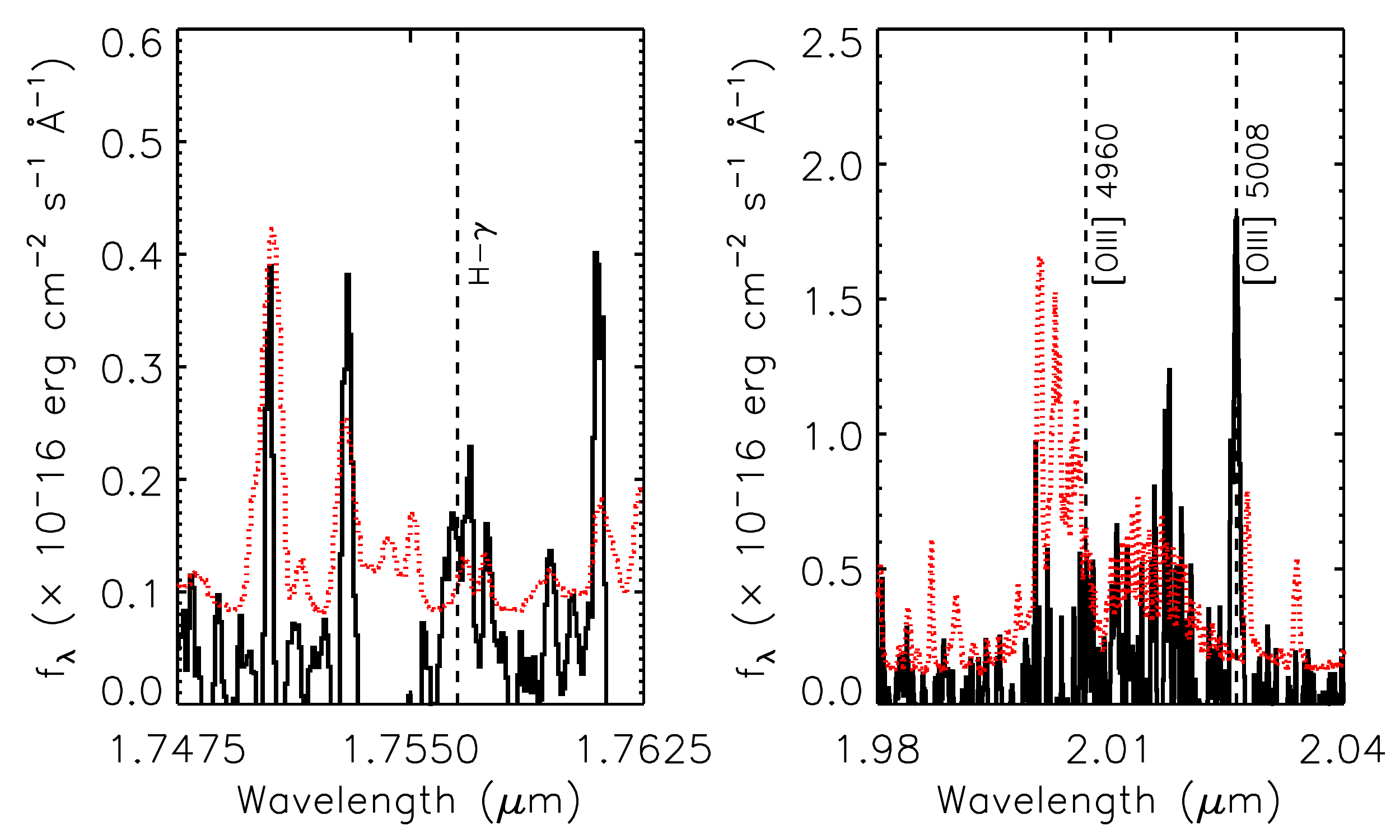}
   \includegraphics[width=0.55\textwidth]{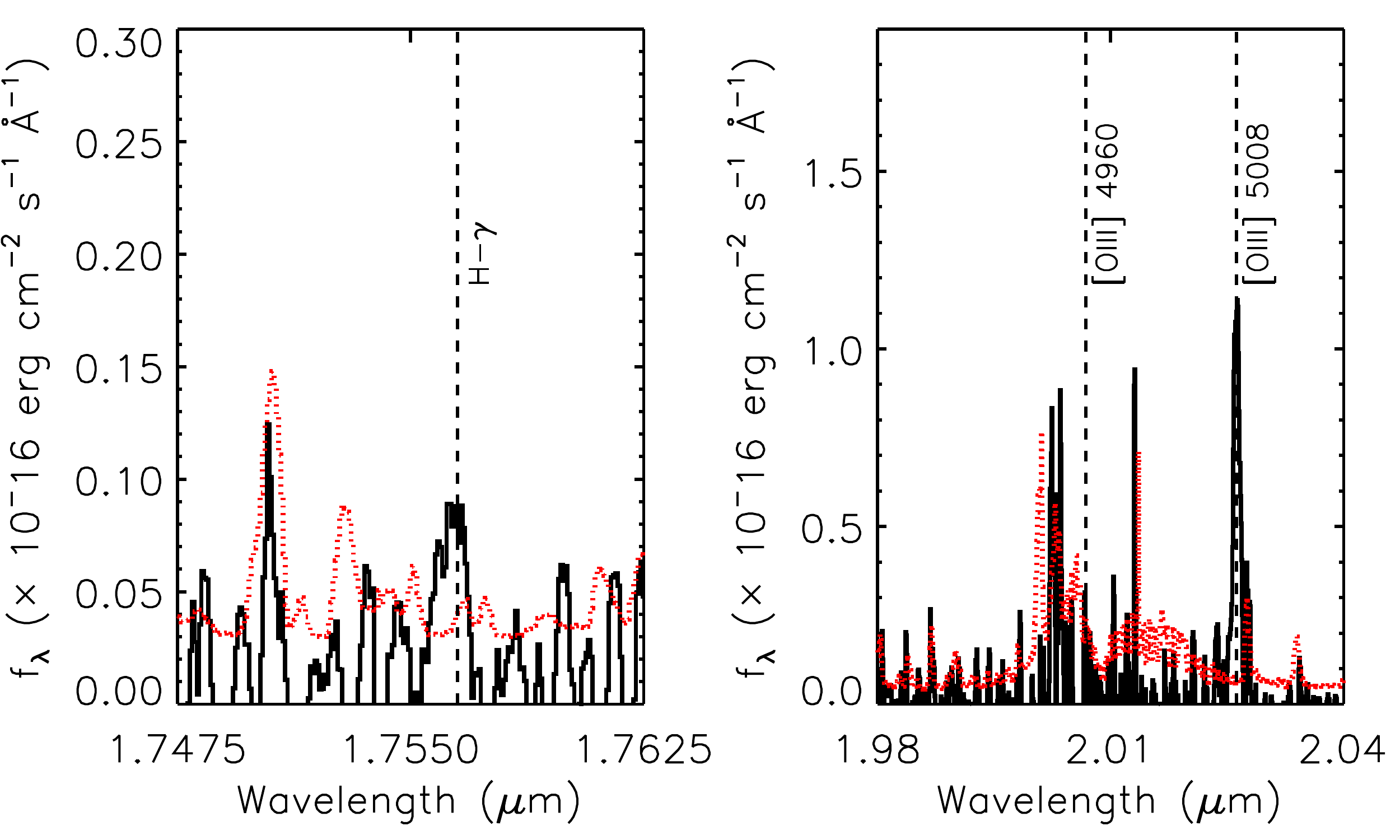}
   \includegraphics[width=0.55\textwidth]{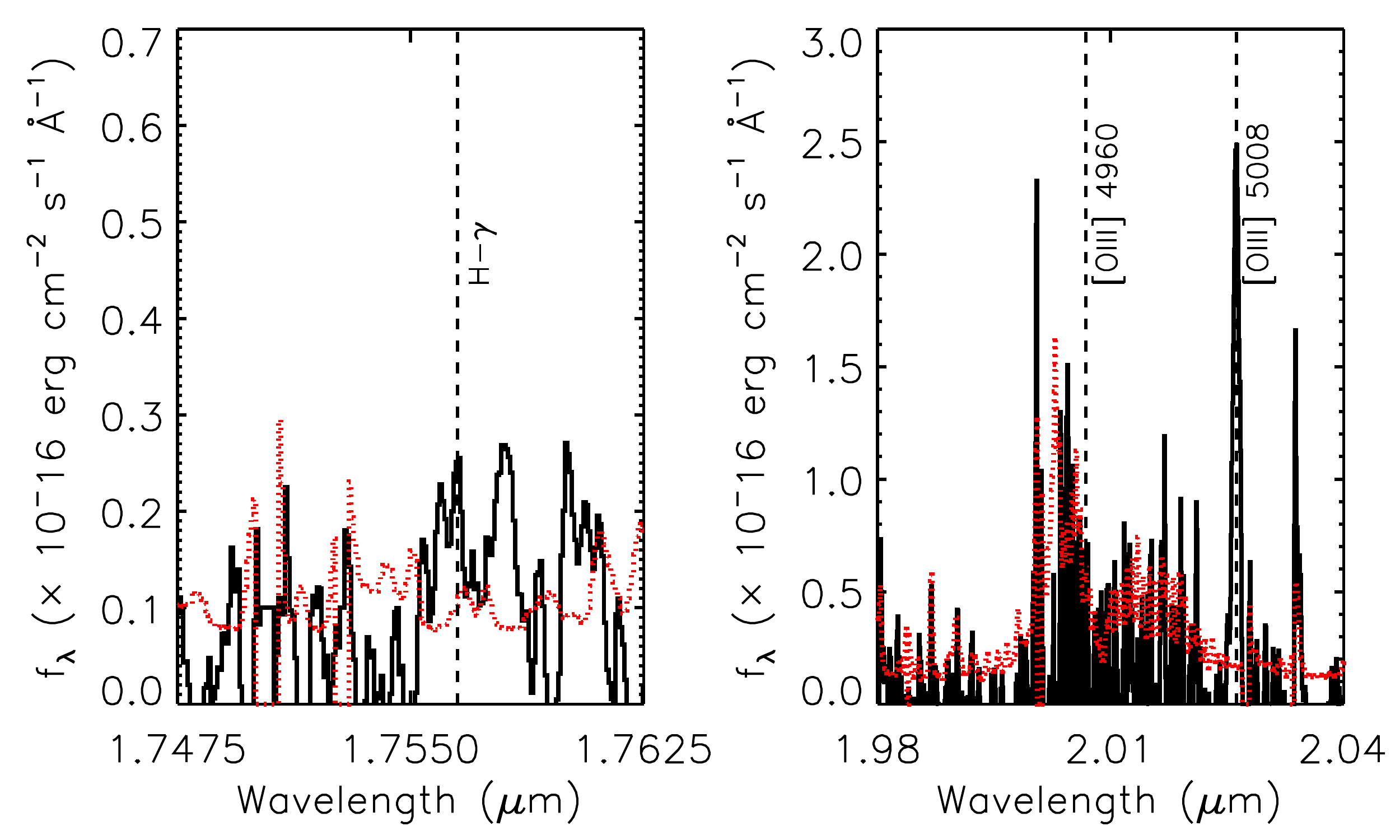}
\end{center}
        \caption{Spectral features identified in the FIRE spectra for three images of system~3  at $z=$ \zspecC. {\it Top:} image~3.1 - {\it middle:} image~3.2 - {\it bottom:} image~3.3.}
    \label{fig:FIREsys3}
\end{figure*}

\end{document}